\documentclass[twoside,twocolumn,9pt]{article}
\usepackage{extsizes}
\usepackage[super,sort&compress,comma]{natbib} 
\usepackage[version=3]{mhchem}
\usepackage[left=1.5cm, right=1.5cm, top=1.785cm, bottom=2.0cm]{geometry}
\usepackage{balance}
\usepackage{mathptmx}
\usepackage{sectsty}
\usepackage{graphicx} 
\usepackage{lastpage}
\usepackage[format=plain,justification=justified,singlelinecheck=false,font={stretch=1.125,small,sf},labelfont=bf,labelsep=space]{caption}
\usepackage{float}
\usepackage{fancyhdr}
\usepackage{fnpos}
\usepackage[english]{babel}
\addto{\captionsenglish}{%
  
}
\usepackage{array}
\usepackage{droidsans}
\usepackage{charter}
\usepackage[T1]{fontenc}
\usepackage[usenames,dvipsnames]{xcolor}
\usepackage{setspace}
\usepackage[compact]{titlesec}
\usepackage{hyperref}

\usepackage{braket}
\usepackage{lipsum}
\usepackage{amsmath}
\usepackage{amssymb}

\usepackage{epstopdf}

\definecolor{cream}{RGB}{222,217,201}

\begin{document}

\pagestyle{fancy}
\thispagestyle{plain}
\fancypagestyle{plain}{
\renewcommand{\headrulewidth}{0pt}
}

\makeFNbottom
\makeatletter
\renewcommand\LARGE{\@setfontsize\LARGE{15pt}{17}}
\renewcommand\Large{\@setfontsize\Large{12pt}{14}}
\renewcommand\large{\@setfontsize\large{10pt}{12}}
\renewcommand\footnotesize{\@setfontsize\footnotesize{7pt}{10}}
\makeatother

\renewcommand{\thefootnote}{\fnsymbol{footnote}}
\renewcommand\footnoterule{\vspace*{1pt}%
\color{cream}\hrule width 3.5in height 0.4pt \color{black}\vspace*{5pt}} 
\setcounter{secnumdepth}{5}

\makeatletter 
\renewcommand\@biblabel[1]{#1}            
\renewcommand\@makefntext[1]%
{\noindent\makebox[0pt][r]{\@thefnmark\,}#1}
\makeatother 
\renewcommand{\figurename}{\small{Fig.}~}
\sectionfont{\sffamily\Large}
\subsectionfont{\normalsize}
\subsubsectionfont{\bf}
\setstretch{1.125} 
\setlength{\skip\footins}{0.8cm}
\setlength{\footnotesep}{0.25cm}
\setlength{\jot}{10pt}
\titlespacing*{\section}{0pt}{4pt}{4pt}
\titlespacing*{\subsection}{0pt}{15pt}{1pt}

\fancyfoot{}
\fancyfoot[LO,RE]{\vspace{-7.1pt}\includegraphics[height=9pt]{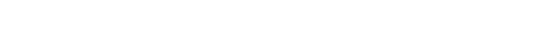}}
\fancyfoot[CO]{\vspace{-7.1pt}\hspace{13.2cm}\includegraphics{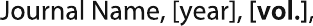}}
\fancyfoot[CE]{\vspace{-7.2pt}\hspace{-14.2cm}\includegraphics{head_foot/RF}}
\fancyfoot[RO]{\footnotesize{\sffamily{1--\pageref{LastPage} ~\textbar  \hspace{2pt}\thepage}}}
\fancyfoot[LE]{\footnotesize{\sffamily{\thepage~\textbar\hspace{3.45cm} 1--\pageref{LastPage}}}}
\fancyhead{}
\renewcommand{\headrulewidth}{0pt} 
\renewcommand{\footrulewidth}{0pt}
\setlength{\arrayrulewidth}{1pt}
\setlength{\columnsep}{6.5mm}
\setlength\bibsep{1pt}

\makeatletter 
\newlength{\figrulesep} 
\setlength{\figrulesep}{0.5\textfloatsep} 

\newcommand{\topfigrule}{\vspace*{-1pt}%
\noindent{\color{cream}\rule[-\figrulesep]{\columnwidth}{1.5pt}} }

\newcommand{\botfigrule}{\vspace*{-2pt}%
\noindent{\color{cream}\rule[\figrulesep]{\columnwidth}{1.5pt}} }

\newcommand{\dblfigrule}{\vspace*{-1pt}%
\noindent{\color{cream}\rule[-\figrulesep]{\textwidth}{1.5pt}} }

\makeatother

\twocolumn[
  \begin{@twocolumnfalse}
{\includegraphics[height=30pt]{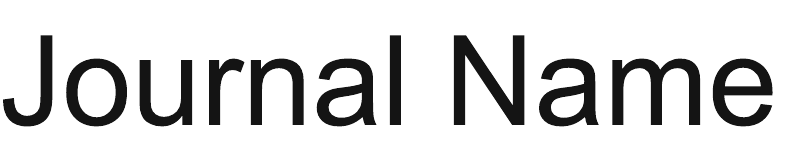}\hfill\raisebox{0pt}[0pt][0pt]{\includegraphics[height=55pt]{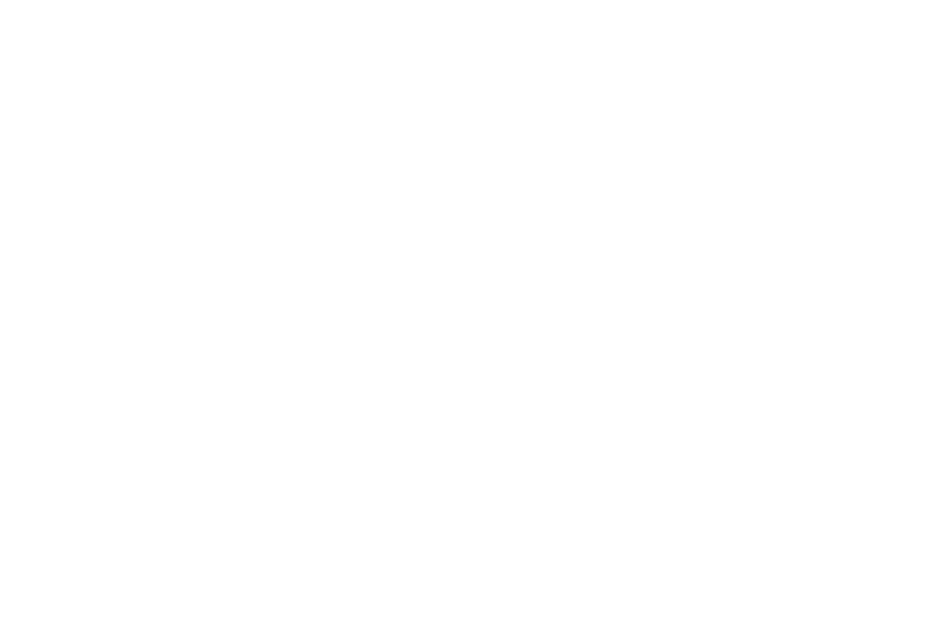}}\\[1ex]
\includegraphics[width=18.5cm]{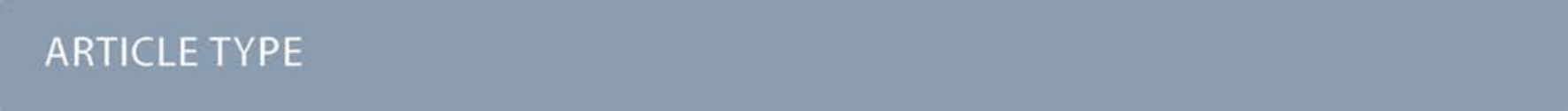}}\par
\vspace{1em}
\sffamily
\begin{tabular}{m{4.5cm} p{13.5cm} }

\includegraphics{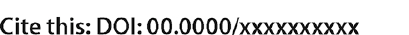} & \noindent\LARGE{\textbf{Broadband cavity-enhanced ultrafast spectroscopy}} \\
\vspace{0.3cm} & \vspace{0.3cm} \\

 & \noindent\large{Myles C. Silfies\textit{$^{a}$}, Grzegorz Kowzan\textit{$^{ab}$}, Neomi Lewis\textit{$^{a}$}, and Thomas K. Allison$^{\ast}$\textit{$^{a}$}} \\

\includegraphics{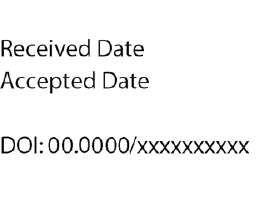} & \noindent\normalsize{Broadband ultrafast optical spectroscopy methods, such as transient absorption spectroscopy and 2D spectroscopy, are widely used to study molecular dynamics. However, these techniques are typically restricted to optically thick samples, such as solids and liquid solutions. In this article we discuss a cavity-enhanced ultrafast transient absorption spectrometer covering almost the entire visible range with a detection limit of $\Delta$OD $ < 1 \times 10^{-9}$, extending broadband all-optical ultrafast spectroscopy techniques to dilute beams of gas-phase molecules and clusters. We describe the technical innovations behind the spectrometer and present transient absorption data on two archetypical molecular systems for excited-state intramolecular proton transfer, 1'-hydroxy-2'-acetonapthone and salicylideneaniline, under jet-cooled and Ar cluster conditions.} \\

\end{tabular}

 \end{@twocolumnfalse} \vspace{0.6cm}

  ]

\newcommand{\units}[1]{\mbox{ }\mbox{#1}}
\newcommand{\ind}{\hspace{0.2 in}}
\newcommand{\textss}[1]{\scriptsize \mbox{#1}}
\newcommand{\vecbf}[1]{\mathbf{#1}}
\newcommand{\abs}[1]{\left| #1 \right|}
\newcommand{\sgn}{\operatorname{sgn}}
\newcommand{\intpminfty}{\int_{-\infty}^{\infty}}
\parindent= 0.0in

\renewcommand*\rmdefault{bch}\normalfont\upshape
\rmfamily
\section*{}
\vspace{-1cm}


\footnotetext{\textit{$^{a}$~Departments of Chemistry and Physics, Stony Brook University, Stony Brook, NY 11790-3400, USA; E-mail: thomas.allison@stonybrook.edu}}
\footnotetext{\textit{$^{b}$~Institute of Physics, Faculty of Physics, Astronomy and Informatics, Nicolaus Copernicus University in Toru\'{n}, ul. Grudziadzka 5, 87-100 Toru\'{n}, Poland}}




The spectra of polyatomic molecules that undergo ultrafast dynamics are inherently broad, due both to the energy-time uncertainty principle and also the large number of degrees of freedom usually involved in the dynamics. Thus, in general the spectral "blobs" observed in the so-called \emph{linear} spectra of polyatomic molecules in the visible and ultraviolet are not particularly informative regarding the underlying dynamics. Ultrafast spectroscopy techniques attempt to address this problem by observing dynamics directly in the time domain. Put another way, by using a \emph{nonlinear} spectroscopy, in which the molecule interacts with multiple photons, one tries to ``parse the blob" into sub-components which may distinguish themselves with different kinetics, orientational dynamics, or spectral correlations in the case of 2D spectroscopy. However, even the parsed blob, for example broken down into constituent parts by global analysis \cite{vanStokkum_GlobalAnalysis2004}, still often leaves much room for interpretation in assigning the components of ultrafast spectra and extracting the relevant physical quantities.
 
\ind Spectral assignments aside, even interpreting the seemingly simplest aspect of ultrafast spectroscopy data---extracting kinetic time constants from the decay of signals with increasing pump/probe delay---is not simple. Although thousands of such ``lifetimes"  are published every year, the measurement does not actually give an excited-state lifetime. Rather, when a time-dependent excited state $\Psi(t)$ is probed in a pump/probe experiment, what is actually recorded are \emph{projections}:

\begin{equation}\label{eqn:simpleprojection}
	S_f(t) \propto \left| \braket{\Psi_f | \hat{\mu} | \Psi(t)} \right|^2\ 
\end{equation}

where the $\Psi_f$ are final states and  $\hat{\mu}$ is the dipole operator that connects the molecule to the electromagnetic field. For example, ``energy windowing" effects can have a large impact on the kinetic time constants observed in time-resolved photoelectron spectroscopy experiments \cite{Hudock_JPCA2007, Tao_JChemPhys2011, Adachi_PCCP2019, Liu_PRX2020}. Even without windowing effects, in general, any ultrafast experiment in polyatomic molecules necessarily projects dynamics with many active degrees of freedom onto an observable with far fewer dimensions, resulting in significant information loss. The nature of the chosen projection can then have a profound impact on interpreting the results.   

\begin{figure}
\centering
  \includegraphics[width=0.8\linewidth]{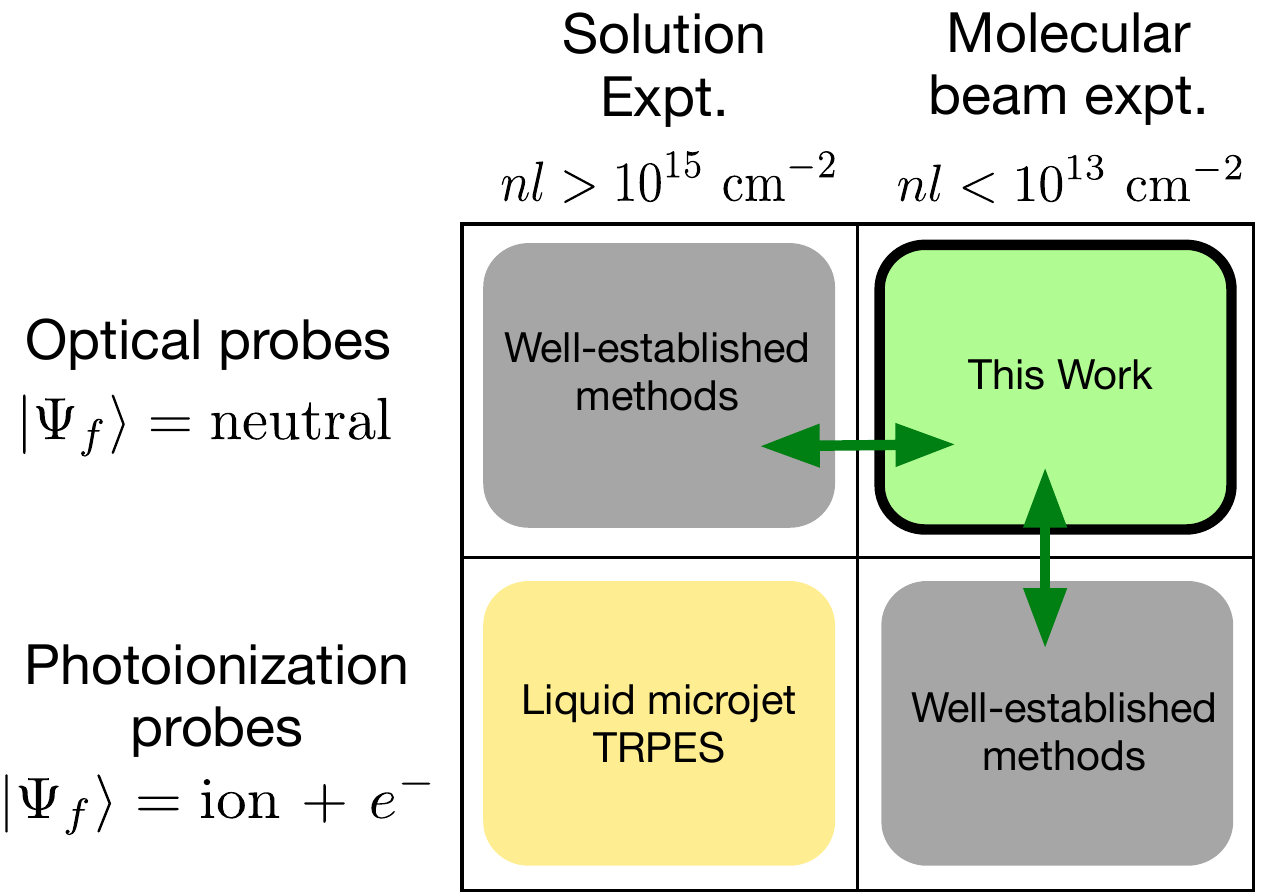}
  \caption{\textbf{Overview of ultrafast spectroscopy methods. By measuring the same observable as most solution-phase ultrafast spectroscopy studies, but on jet-cooled molecules and clusters, cavity-enhanced ultrafast transient absorption spectroscopy (this work) establishes a link between gas-phase methods based on photoionization and solution-phase ultrafast dynamics studies based on optical probes.}}
  \label{fig:punnett}
\end{figure}

\ind Ideally, to get maximum information, one would project the state of interest $\Psi(t)$ onto as many final states as possible and make comparisons between systems prepared differently using the same observables. It has also recently emerged that it is critical to compare the experimental signals with theory that directly simulates the experiment by calculating the relevant observables, performing the same projections (Eq.~\eqref{eqn:simpleprojection}) \emph{in silico} that are done by the experiments in the lab \cite{Hudock_JPCA2007, Tao_JChemPhys2011, Allison_JChemPhys2012, Schuurman_AnnRevPhysChem2018, Liu_PRX2020}. Comparisons with \emph{ab initio} theory are most robustly done for gas-phase systems, but the bulk of ultrafast spectroscopy is done in solutions, and almost always with very different observables than gas-phase studies. The current paradigm is illustrated in figure \ref{fig:punnett}. Optical methods, such as transient absorption (TAS) and 2D spectroscopy, are well-established for solution-phase work ($\Psi_f = $ neutral states). In contrast, gas-phase experiments, particularly in the physical chemist's playground of molecular beams, rely almost exclusively on photionization methods such as time-resolved photoelectron spectroscopy (TRPES) \cite{Stolow_ChemRev2004} -- action spectroscopies that project the system onto the 1-electron continuum ($\Psi_f = $ free electron + cation).

\ind Comparing results using these different observables can be very difficult. A good example is the problem of internal conversion processes in nucleo-bases, responsible for the UV photo-protection of DNA \cite{Kohler_AnnRevPhysChem2009, Saigusa_JPhotoChemBio2007}. Researchers studying isolated gas-phase molecules using TRPES \cite{Saigusa_JPhotoChemBio2007, Hudock_JPCA2007, Ullrich_PCCP2004} report quite different dynamics than those studying solution-phase molecules with TAS \cite{Kohler_AnnRevPhysChem2009} or 2D spectroscopy \cite{Tseng_JPCA2012}. Undoubtedly, the dynamics are different in solution, but the very different observables and disparate data sets obtained with the different experimental techniques cloud the comparison, to the point where the practitioners of the different measurements almost form separate communities with their own separate review papers \cite{Kohler_AnnRevPhysChem2009, Saigusa_JPhotoChemBio2007}.   

\ind  In a previous article, Reber, Chen, and Allison \cite{Reber_Optica2016} described the extension of ultrasensitive direct absorption techniques to femtosecond time-resolved experiments, reporting cavity-enhanced all-optical measurements in a dilute molecular beam that are simultaneously ultrasensitive and ultrafast. Using frequency combs and optical resonators, cavity-enhanced transient absorption (CE-TA), or pump-probe, measurements were demonstrated with a time resolution of 120 fs and a detection limit for changes in sample absorbance of $\Delta $OD $ = 2 \times 10^{-10}$, an improvement over the previous state of the art \cite{Schriever_RSI2008} by nearly four orders of magnitude. This large advance in sensitivity can enable many measurements previously thought impossible. However, this previous demonstration operated at only one wavelength (530 nm). One wavelength does not make a spectrum, and the inherently broad spectra of chemically-relevant molecules undergoing ultrafast dynamics demand wide spectral coverage.

\ind In this article, we report the development of a broadband cavity-enhanced ultrafast transient absorption spectrometer (CE-TA\textbf{S}) operating across the wavelength band of 450-700 nm---a bandwidth greater than 7900 cm$^{-1}$ (240 THz) covering almost the entire visible spectral range. To go from CE-TA to CE-TAS has involved considerable innovation, since many of the necessary components did not exist prior to our work. We have previously published results regarding aspects of the optical technology critical to CE-TAS, namely the development of widely tunable, low noise, high-power frequency combs \cite{Chen_ApplPhysB2019} and the enhancement of these widely-tunable combs in a femtosecond enhancement cavity with custom mirror coatings \cite{Silfies_OptLett2020}. Achieving reliable and reproducible transient absorption spectroscopy data with $\Delta$ OD $ < 10^{-9}$ using this optical technology has also required significant innovation which we detail in this paper, where we present the first spectroscopy results from this system. To our knowledge, this also the first cavity-enhanced comb spectroscopy of any kind (ultrafast or otherwise) using a widely tunable platform.

\ind This work establishes CE-TA\textbf{S} as a new broadly applicable technique for gas-phase chemical physics, and creates a missing link between gas-phase and solution-phase studies shown in figure \ref{fig:punnett}. For gas-phase molecules, UV-visible CE-TAS provides another projection of the dynamics complimentary to gas-phase TRPES, with a dataset that is directly comparable to solution-phase work via the common observable. Cluster studies enabled by CE-TAS also allow probing intermediate levels of solvation. We note that others pursue a similar linking path via attempting TRPES on molecules in solution via the liquid micro-jet approach \cite{Suzuki_JChemPhys2019}, as also illustrated in figure \ref{fig:punnett}. Sensitivity is also the challenge in these experiments to move beyond neat liquids or ultra-concentrated solutions to pump/probe experiments on more chemically relevant systems\cite{Suzuki_JChemPhys2019}. 

\ind In addition to filling the gap illustrated in figure \ref{fig:punnett}, we believe these broadband CE-TAS methods can also be adapted for work on solids, ultra-dilute solutions, or sparsely covered surfaces that would benefit from improved sensitivity. In the sections below we describe the many unique aspects of the spectrometer and a detailed analysis of its performance.

\section{Experimental Setup}

\subsection{Light Sources and Enhancement Cavity}

\begin{figure*}[h!]
		\centering
		\includegraphics[width = 0.8\linewidth]{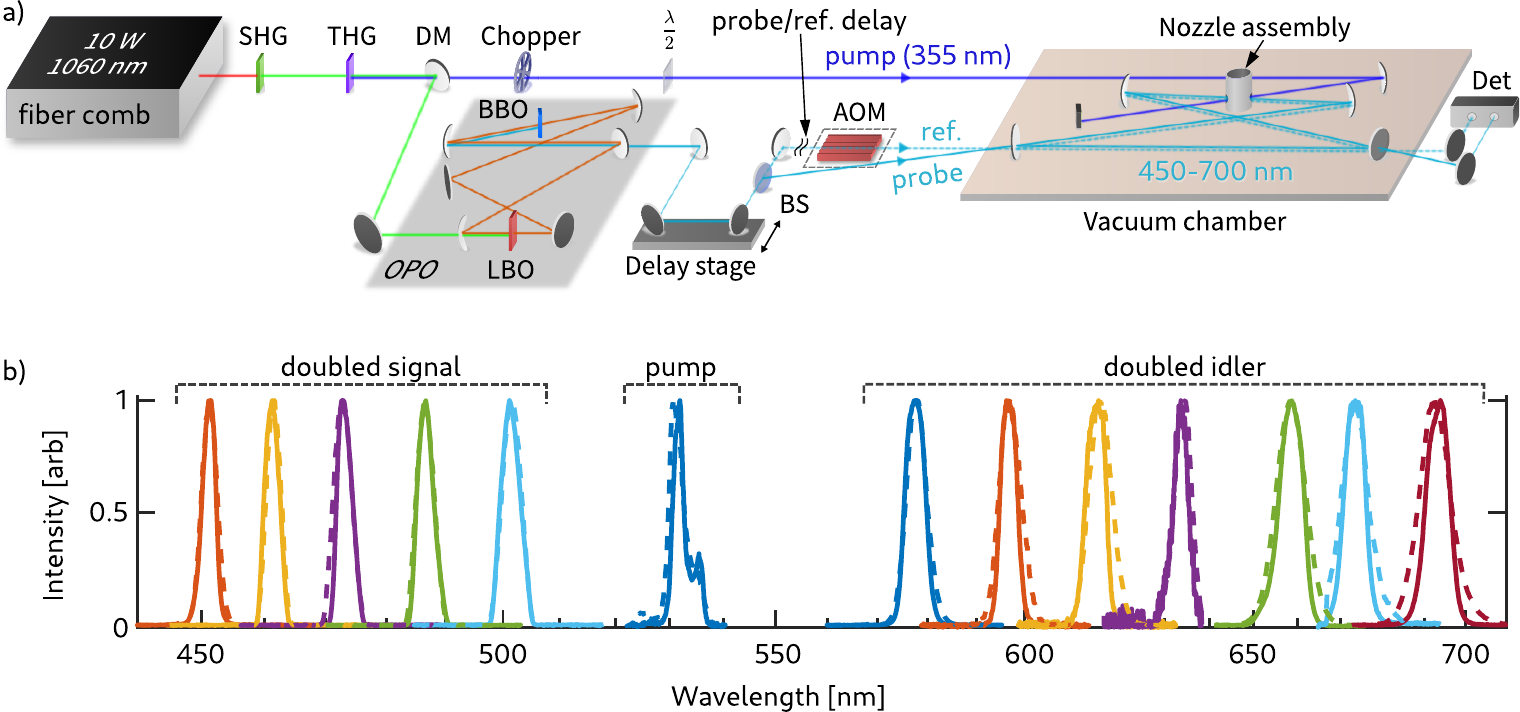}
		\caption{a) Optical layout of the broadband cavity-enhanced transient absorption spectrometer. Tunable frequency combs are derived from a synchronously pumped optical parametric oscillator (OPO) and coupled to a 4-mirror broadband dispersion-managed enhancement cavity. The third harmonic of the Yb:fiber comb at 355 nm is used for molecule excitation in the current experiments. More details regarding the optical components are in the main text and references \citenum{Silfies_OptLett2020, Chen_ApplPhysB2019}. b) OPO (dashed) and cavity-enhanced (solid) spectra across the OPO 450-700 nm tuning range. Broadband spectra are assembled from pump/probe traces recorded with different OPO wavelengths.}
		\label{fig:opt_setup}	
\end{figure*}

The optical setup is illustrated in figure \ref{fig:opt_setup}a). We derive the initial frequency comb at 1064 nm from a 1550 nm Er:fiber oscillator (Menlo Systems Ultra-Low-Noise variant), shifted to 1064 nm using dispersive-wave generation in a short highly nonlinear fiber \cite{Maser_ApplPhysB2017}. We then amplify the shifted Er:fiber comb to 10 W average power in a home-built large-mode-area Yb-doped photonic crystal fiber amplifier previously described \cite{Li_RSI2016}. The  100 MHz repetition-rate ($f_{\textss{rep}}$) amplified pulse train from this laser is frequency doubled and tripled (2+1) in critically phase-matched lithium trioborate (LBO) and beta barium borate (BBO) crystals, respectively. We use the third harmonic at 355 nm from this setup, with approximately 500 mW of average power, for the pump in the CE-TAS measurements presented here. In the present measurements, working with molecules with relatively large excitation cross sections, we obtain sufficient signal to noise without employing an enhancement cavity for the pump to boost the pump power, as was done in Reber et al. \cite{Reber_Optica2016}, but it would be relatively straightforward to implement a pump enhancement cavity if even higher sensitivity were needed. We use the residual second harmonic (4.5 W) to pump a home-built tunable synchronously-pumped optical parametric oscillator (OPO) with subsequent intracavity doubling for both the signal and idler \cite{Chen_ApplPhysB2019}. Using the 532 nm pump as well to cover the gap near OPO degeneracy, this provides tunable combs over the range of 420-720 nm, as described in ref. \citenum{Chen_ApplPhysB2019}.

\ind We couple the tunable combs from the OPO to a broadband enhancement cavity with custom mirror coatings optimized to manage group-delay dispersion (GDD) over a wide tuning range, as described in detail in ref. \citenum{Silfies_OptLett2020}. Figure \ref{fig:opt_setup}b) shows representative OPO and enhanced intracavity spectra across the tuning range. The intracavity spectrum are narrower that the OPO spectra due to residual GDD of the enhancement cavity. This sets the limit to the simultaneous intracavity bandwidth, and thus intracavity pulse duration, that can be attained irrespective of the incident comb bandwidth \cite{Jones_OptLett2004,Jones_OptLett2002}. The cavity is in a bow-tie configuration with two plane mirrors for the input and output couplers (nominal 0.3\% transmission), and two high reflectors with 50 cm radius of curvature. With most of the cavity loss coming from the input and output couplers, the cavity is close to the impedance-matched condition \cite{Nagourney_Book2010}. We calculate the beam size ($1/e^2$ radius) to be $w_{\textss{probe}}$ = 65 $\mu$m at 532 nm using the ABCD matrix formalism, and this only scales weakly with probe wavelength as $w_{\textss{probe}} \propto \sqrt{\lambda}$ \cite{Siegman_book1986}. The cavity has a nominal finesse ($\mathcal{F}$) varying from 600 to 1400 across the range of 450-700 nm. OPO output wavelengths outside this range are not used due to the limits of the cavity mirror high-reflectance band. We focus the 355 nm pump beam to a waist size of approximately $w_{\textss{pump}}$ =150 $\mu$m and overlap the pump focus with the enhancement cavity focus above the molecular beam source, as illustrated in figure \ref{fig:opt_setup}. The pump beam is chopped at a frequency between 3 and 4 kHz, well inside the enhancement cavity's minimum linewidth of 70 kHz (above which the cavity would low-pass filter the CE-TA signal unless higher-order modes are used \cite{Nagourney_Book2010, Allison_JPhysB2017}), but above the lab's $1/f$ noise.

\ind Although the residual OPO pump (532 nm), doubled signal ($2s$), and doubled idler ($2i$) combs follow the same optical path, there are substantial differences to the setup for using each of these three combs. First of all, the OPO optical-phase transfer relations we discovered in ref.\citenum{Chen_ApplPhysB2019} necessitate that the three different combs are frequency-locked to the enhancement cavity using three different schemes with different actuators, as detailed in ref. \citenum{Silfies_OptLett2020}. Furthermore, there are substantial differences in the relative intensity noise (RIN) spectra of the intracavity light before the common-mode noise rejection scheme described below is applied\cite{Silfies_OptLett2020}. Also, we change the mode-matching optics between the OPO and the enhancement cavity when changing between output combs to account for different spatial modes and divergence from the OPO. Despite all these differences, comparable CE-TAS performance can be obtained using all three combs as we show in section \ref{sec:results}. 

\subsection{Vacuum System and Supersonic Expansion}

The enhancement cavity is mounted on a 60 cm $\times$ 120 cm breadboard inside a rectangular vacuum chamber. The breadboard is supported via legs that protrude through the bottom of the chamber via bellows down to the optical table. In this way the breadboard is isolated from vibrations of the vacuum chamber or flexure of the vacuum chamber upon pump out. 

 \ind Molecules are introduced at the common focus of the probe cavity and the pump beam using a continuously-operating slit nozzle. A planar expansion, as opposed to an axisymmetric expansion from a pinhole, is used to attain a higher column density of molecules and also facilitate cluster studies \cite{Miller:1988}. The gas load of the continuous planar expansion is handled by a three-stage pumping system consisting of two Roots pumps (5000 m$^3$/hr and 1400 m$^3$/hr) in series backed by a two-stage 100 m$^3$/hr oil-sealed rotary vane pump. 
  
\begin{figure}[t!]
	\centering
	\includegraphics[width=\linewidth]{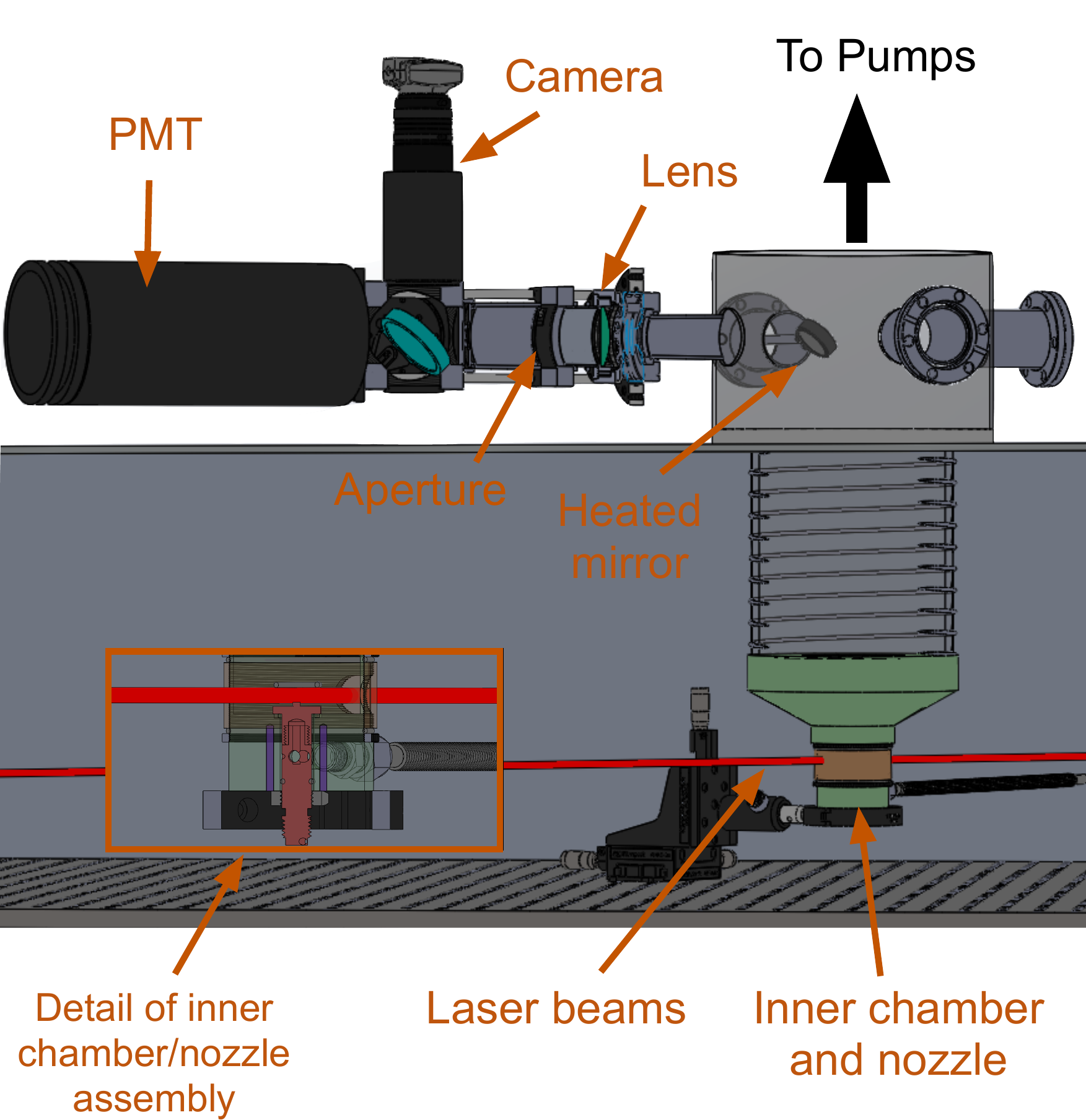}
	\caption{Molecular beam setup and fluorescence monitor. The fluorescence detection scheme is described in detail in the main text. Inset shows a cutaway of the nozzle assembly. The inner chamber surrounding the heated nozzle contains the sample molecule near the pump/probe overlap region.}
	\label{fig:gas_setup}
\end{figure}

 \ind To prevent cavity mirror contamination, the supersonic expansion takes place in a small inner chamber inside the main vacuum chamber, as shown in figure \ref{fig:gas_setup}. The inner chamber is maintained at $\sim$100 mTorr via the Roots pumping system. The inner chamber is connected to the main chamber via two 3 mm holes that allow the laser beams to pass through. We then flow Argon gas into main chamber which creates a flow of Ar into the inner chamber via these 3 mm holes. This steady purging flow prevents sample molecules from exiting the inner chamber. Argon is used instead of nitrogen to avoid possible artifacts due to non-resonantly excited rotational coherences \cite{Jarzeba_PCCP2002}. Typical argon pressures in the main chamber are $\sim$ 10 mbar, which is sufficient to prevent mirror contamination, but small enough that it does not produce enough group delay dispersion to narrow the enhanced comb bandwidth \cite{Pupeza_OptExp2010}. A small flow of oxygen is also directed at each cavity mirror to further help mitigate hydrocarbon contamination.

 \ind For introducing non-volatile molecules to the experiment, molecules are sublimed at temperatures up to 150$^{\circ}$ C in a cell external to the vacuum chamber and then entrained in a flow of noble carrier gas. The supersonic nozzle assembly and associated gas feedline are also heated to prevent molecule condensation. Typical sample consumption rates are 0.5-3 g/hr. 

\subsection{Procedure for Individual CE-TA Pump/Probe Trace Accumulation}

\begin{figure}[t!]
\centering
  \includegraphics[width=\linewidth]{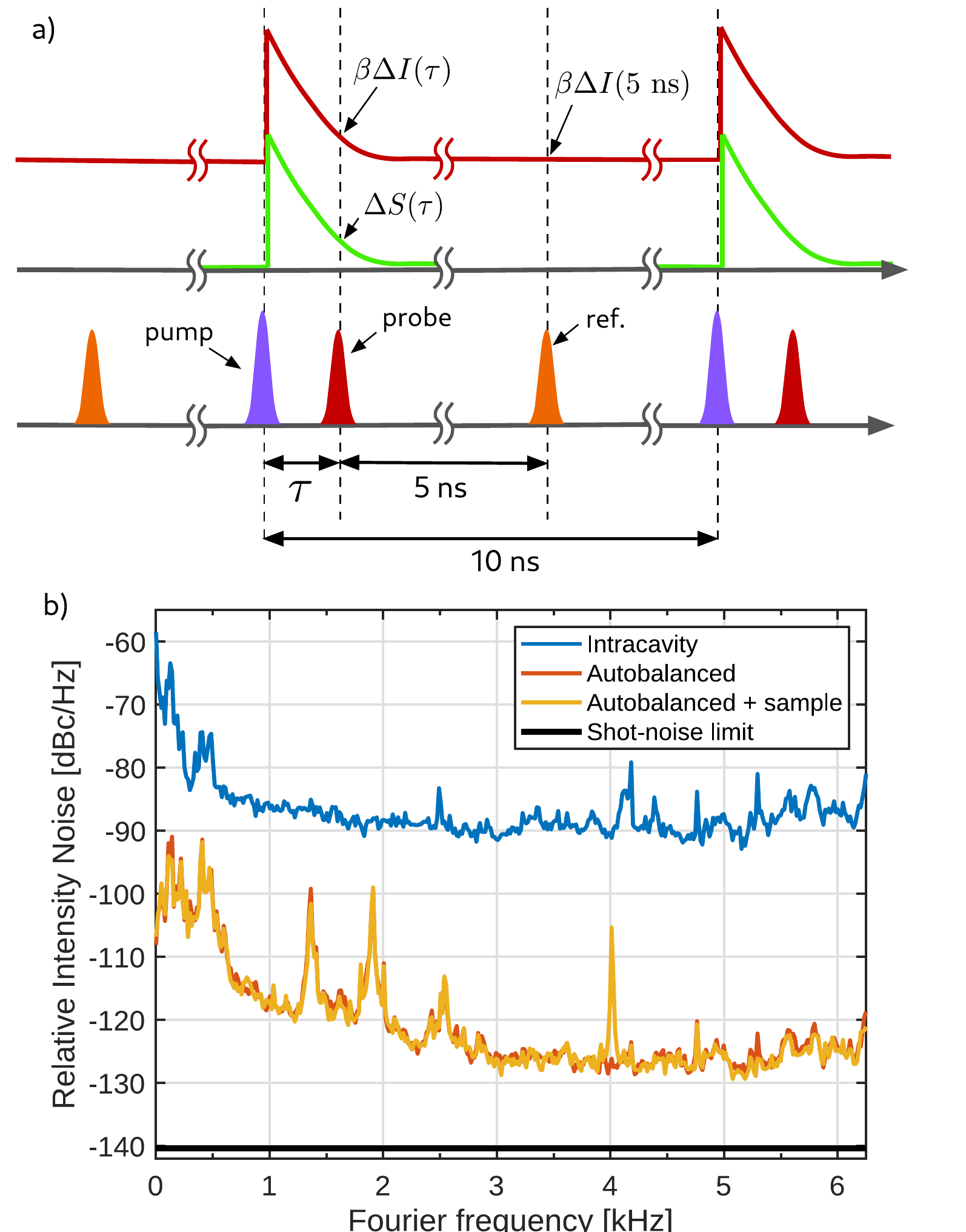}
  \caption{a) Pulse sequence at the sample. The reference pulse records any steady-state pump/probe signal $\Delta I(5\units{ns})$ and contains nearly identical noise to the probe for common mode subtraction.  b) Noise spectrum of the intracavity light and subtracted signal using 469 nm (2$s$) light. Rejection of common-mode noise using the autobalanced subtraction scheme allows for the ultrafast molecular signal to be detected at the pump modulation frequency of 4 kHz. Also shown is the shot-noise (or quantum noise) limit calculated from the measured photocurrent.}
  \label{fig:subtract}
\end{figure}

When recording transient absorption measurements using any of the three combs, we couple a delayed reference pulse train to the cavity in a counter-propagating direction as shown in figure \ref{fig:opt_setup}a). The resulting pulse sequence at the molecular beam is shown in figure \ref{fig:subtract}a). The reference beam pulses arrive $\sim$5 ns later than the probe and pump. The normalized pump/probe CE-TA signal $(\Delta S)$ at each OPO wavelength is recovered via autobalanced subtraction (probe$-$reference)\cite{Hobbs_AppOpt1997} and lock-in detection at the pump modulation frequency, such that the CE-TAS signal is given by

\begin{equation}\label{eqn:dS}
	\Delta S(\tau) = \frac{\pi}{\mathcal{F}} \frac{\Delta I (\tau) - \Delta I(\tau + 5 \mbox{ ns})} {I_{\textss{probe}}} \equiv \beta \left[ \Delta I (\tau) - \Delta I(\tau + 5 \mbox{ ns}) \right]
\end{equation}

where $\tau$ is the pump/probe delay, $I_{\textss{probe}}$ is the intracavity light intensity for the probe beam, the $\Delta I$ are pump-induced changes in the intracavity light intensity, and the factor $\pi/\mathcal{F}$ is the inverse of the cavity enhancement cavity for impedance-matched cavity and our experimental geometry \cite{Gagliardi_Book2013}. The subtraction accomplishes two critical tasks. First and most important is common-mode noise subtraction. The probe ($\Delta I (\tau)$) and reference ($\Delta I(\tau + 5 \mbox{ ns}) \approx \Delta I(5 \mbox{ ns}) $) share mostly the same noise, but have different pump/probe delay-dependent signals due to their timing with respect to the pump pulse train. Since at $\tau$ + 5 ns all fast dynamics have subsided, the subtraction retrieves the femtosecond-delay dependent signal from the noise. Figure \ref{fig:subtract}b) shows the effect of this common-mode noise rejection scheme on the relative intensity noise (RIN) of the intracavity light. With autobalanced subtraction, the noise floor of the measurement is within one order of magnitude (20 dB in RIN) of the quantum noise limit. 

\ind Second is that the $\Delta I(\tau + 5 \mbox{ ns})$ reference signal also contains any signal due to repetitive pumping of the sample or molecular excitation that lives longer than $1/f_{\textss{rep}}$ = 10 ns. Another way to think of this is that due to the 100 MHz repetition rate, in steady state $\Delta I(\tau = 5 \mbox{ ns})$ = $\Delta I(\tau = -5 \mbox{ ns})$ such that the subtraction of the reference signal removes any signal due to preceding pump/probe pulse sequences. This is relevant since for a molecular beam speed of 500 m/s (e.g. for an Ar supersonic expansion), each molecule sees approximately $f_{\textss{rep}} \times (300\text{ }\mu\text{m}/500 \units{m/s}) = 60$ pump pulses. The problem can be exacerbated via velocity slip between the sample molecule and the carrier gas, and even for molecules with short-lived excited states, a ground-state bleach signal may persist. Subtraction of any persistent signal enables CE-TAS to work even with these complications. For most purposes $\Delta S(\tau)$ can be regarded as the femtosecond to picosecond component of the true TAS signal induced by a single pump pulse, simply with a DC offset subtracted. However, one must be aware of subtleties. For example, since the absolute signal size is reduced via subtraction of $\Delta I(5 \mbox{ ns})$, care must be taken in considering signal ratios as discussed in section \ref{sec:results}.

\ind The polarization of the probe light is horizontal ($p$). The pump polarization is controlled to be either $p$ or $s$ (vertical) with a zero-order half wave plate to give pump/probe signals for both parallel ($\Delta S_{\parallel}$)  and perpendicular ($\Delta S_{\perp}$) polarization conditions, respectively. We construct magic-angle signals, insensitive to molecular orientation or rotational coherences, via $\Delta S_{MA} = (\Delta S_{\parallel} + 2\Delta S_{\perp})/3$ \cite{Weiner_book2009, Felker_JChemPhys1987}. Another interesting subtlety of CE-TAS is that magic-angle data cannot be recorded simply by orienting the pump polarization 54.7$^{\circ}$ to the probe, as is usually done in transient absorption spectroscopy. This is due to the fact that the non-zero angles of incidence on the enhancement cavity mirrors causes the $p$ and $s$ eigenmodes of the cavity to be non-degenerate. Thus light scattered into an $s$ mode of the cavity by a magic angle pump would not be exactly on resonance, leading to increased noise and also a different signal enhancement. Using only $s$ and $p$ pump polarizations ensures that the intracavity probe light remains $p$-polarized by symmetry.

\subsection{Constructing Transient Absorption Spectra}
\label{sec:Construct}
The probe bandwidth of each individual CE-TA measurement describe above, with the OPO output tuned to a particular wavelength, is less than 10 THz (figure \ref{fig:opt_setup}b). We thus assemble broadband transient absorption spectra by combining a collection of measurements taken at different wavelengths. Controlling systematics is then of the utmost importance to assemble reliable and reproducible CE-TAS spectra, as several parameters affecting the signal size vary intrinsically with wavelength and can also vary with time over the course of an experimental run.

\ind To control for cavity finesse variation, we periodically perform in-situ cavity ring-down measurements at each wavelength in between pump/probe delay scans. We do this by inserting an acousto-optic modulator (AOM) in the reference beam to quickly ($\sim 20$ ns) turn off the reference beam while the Pound-Drever-Hall lock between the comb and cavity is maintained using the probe beam. To achieve 100\% turn-off of the reference beam for clean ring-down signals, we use the first-order diffracted beam from the AOM. The AOM is driven by a 2$f_{\textss{rep}}$ radio-frequency signal derived from the Er:fiber comb. Using an integer multiple of $f_{\textss{rep}}$ to drive the AOM ensures that the frequency-shifted diffracted comb is still resonant with the enhancement cavity. 

\ind To control for potential variations in pump power and sample molecule density at the focus, we record fluorescence from the pump/probe interaction region using the scheme shown in figure \ref{fig:gas_setup}. A mirror in the supersonic expansion path reflects fluorescence out of the chamber. The mirror is heated to prevent sample molecule condensation. To eliminate scattered light background, we then use an f = 10 cm lens to image the pump/probe overlap region to an adjustable aperture which rejects light from elsewhere. The remaining light from the pump/probe overlap region is recorded with a photomultiplier tube (PMT) using lock-in detection at the pump modulation (chopper) frequency. This scheme produces a background-free fluorescence signal that is proportional to the column density of excited molecules in the focal region, and this fluorescence signal can then used to normalize and combine pump/probe data accumulated over extended periods of time.  
 
\section{Results} \label{sec:results}

\subsection{Individual CE-TA Measurements}

\begin{figure}[b!]
	\centering
	\includegraphics[width = 0.7\linewidth]{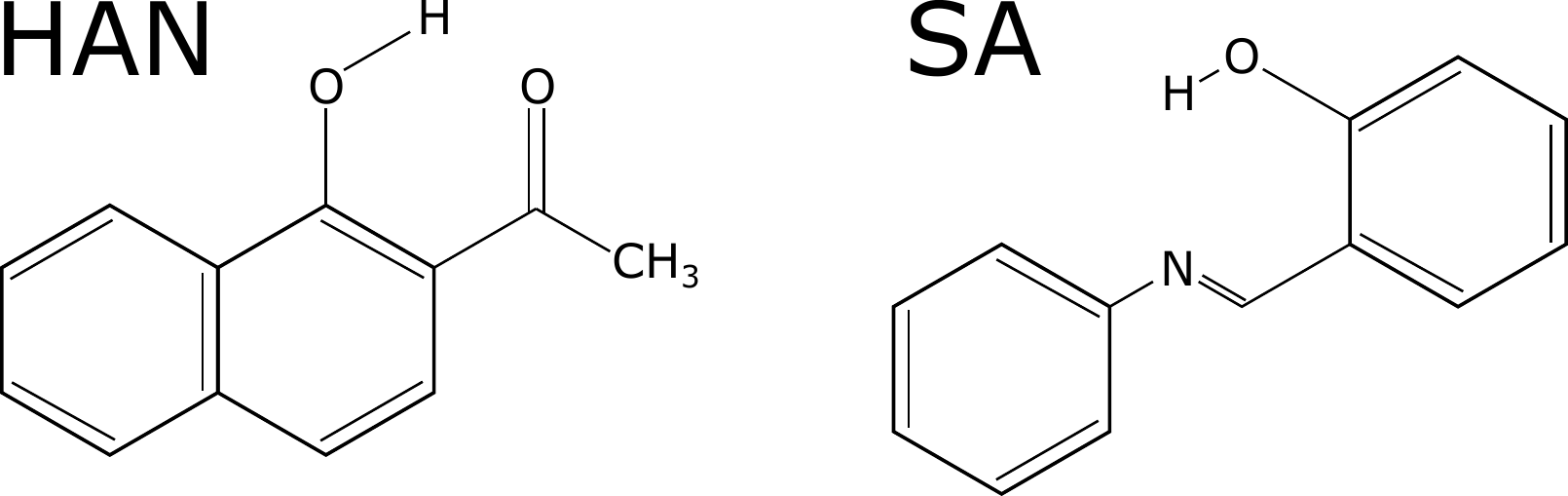}
	\caption{Molecules in the present experiments. HAN = 1'-hydroxy-2'-acetonapthone. SA = salicylideneaniline}
	\label{fig:molecules}
\end{figure}

For the present demonstration of the instrument, we present results on 1'-hydroxy-2'-acetonaphthone (HAN), and Salicylideneaniline (SA), two archetypal systems for excited-state intramolecular proton transfer (ESIPT) shown in figure \ref{fig:molecules}. These molecules have previously been studied using both solution-phase TAS \cite{Lochbrunner_JChemPhys2005, Ziolek_PCCP2004} and gas-phase TRPES \cite{Lochbrunner_JChemPhys2001, Sekikawa_JPCA2013}, so they serve as good systems to benchmark the instrument.

\begin{figure}[t!]
	\centering
	\includegraphics[width=\linewidth]{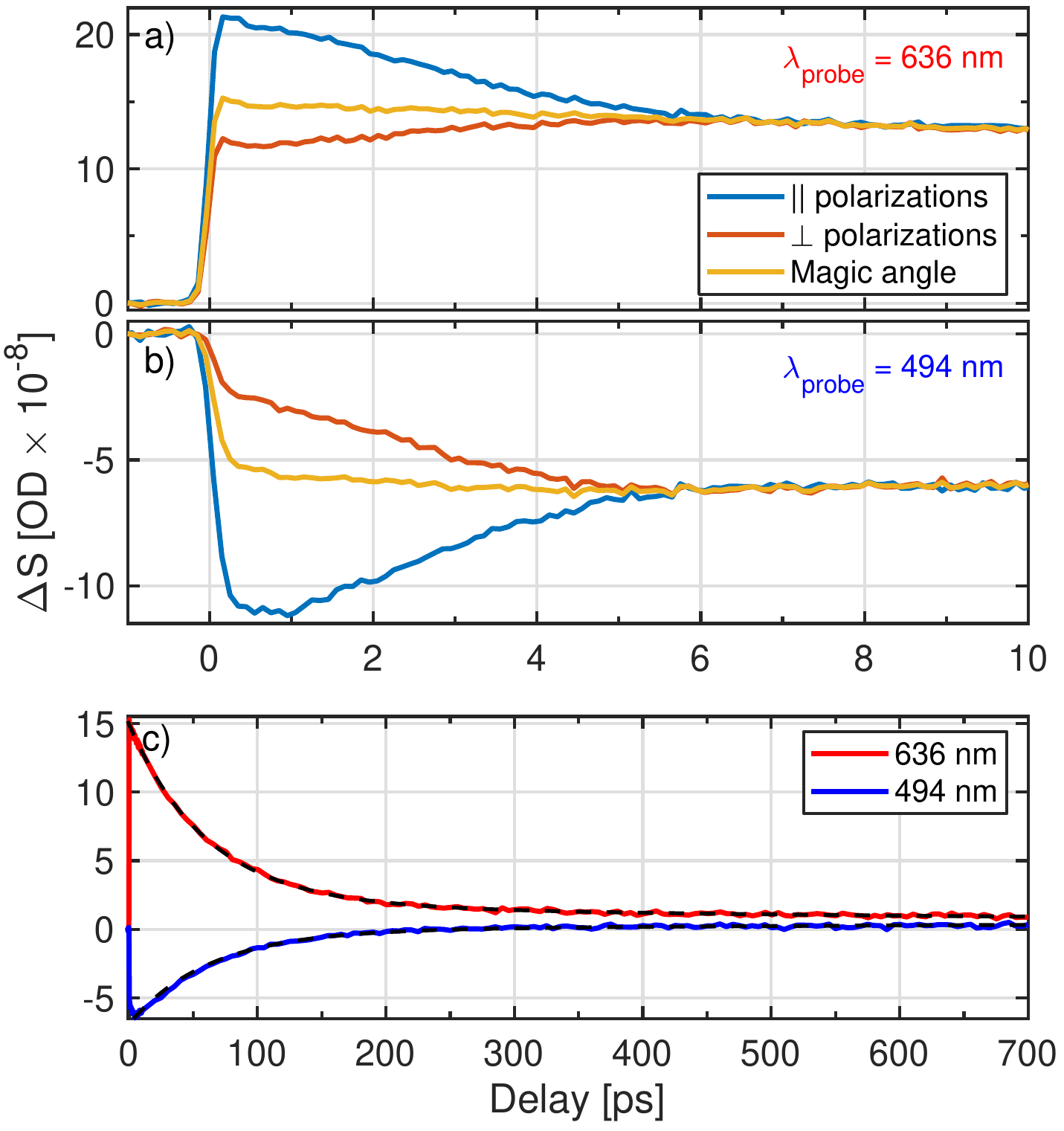}
	\caption{Example transient absorption traces for $\lambda_{\textss{probe}}$ = 636 nm (a) and 494 nm(b) combs recorded from HAN excited at 355 nm. The positive signal in a) corresponds to excited-state absorption and the negative signal in b) to stimulated emission. The parallel and perpendicular polarization data are each the average of three scans with 1s integration time per pump/probe delay. Magic angle data is constructed via $\Delta S_{\textss{MA}} = (\Delta S_{\parallel} + 2\Delta S_{\perp})/3$. c) Full 700 ps magic angle data showing the long decay of the transient signal including single-exponential fits (dashed black lines).}
	\label{fig:TAS_MA}
\end{figure}

\ind Figure \ref{fig:TAS_MA}a) and b) show typical pump/probe data recorded in HAN using the $2i$ and $2s$ combs respectively. Each trace is the average of three scans. Near time zero, a large polarization anisotropy is seen, but this rapidly decays as the many rotational coherences excited by the pump pulse dephase from each other \cite{Felker_JChemPhys1987} in this asymmetric top molecule. Thinking about the problem classically (which is also appropriate here given the large number of rotational states involved) one can think that the pump pulse preferentially excites molecules with their transition dipole oriented along the pump polarization, but then these molecules freely rotate in random directions leading to an isotropic distribution. We note that if one attempts to use the CE-TAS signal (equation \ref{eqn:dS}) to calculate a normalized anisotropy parameter $r'(\tau) = (\Delta S_{\parallel} - \Delta S_{\perp})/(\Delta S_{\parallel} + 2\Delta S_{\perp}) = (\Delta S_{\parallel} - \Delta S_{\perp})/(3 \Delta S_{\textss{MA}})$, this parameter need not be bounded in the usual range of $(-0.2, 0.4)$ due to the subtraction of the long-lived TAS signal sampled by the reference beam. The numerator (a simple difference) gives no artifact, but even if the anisotropy decays to zero at long delays, the denominator of the expression for $r'(t)$ is still reduced by $3\beta\Delta I_{\textss{MA}}$, throwing off the ratio. This is particularly acute for a molecule like HAN, with a fluorescence yield of approximately 1\% and a radiative decay rate of 1/(10 ns) \cite{Catalan_JACS1993}, for which the steady-state excited state population in the focal volume can build up over multiple pump pulses. Indeed, $r'(\tau = 0)$ for the data shown in figure \ref{fig:TAS_MA}b) is 0.6, suggesting a steady-state magic angle background signal of $\Delta I_{\textss{MA}}(5 \units{ns}) = 0.3\Delta I_{\textss{MA}}(\tau = 0)$, which is reasonable under our experimental conditions.

\ind Figure \ref{fig:TAS_MA}c) shows magic-angle pump/probe traces for HAN over the full 700 ps delay range accessible with our delay stage. Fitting these data with a single exponential + offset gives a time constant of 70 ps for internal conversion in HAN, in agreement with previous TAS measurements in cyclohexane \cite{Lochbrunner_JChemPhys2005} and fluorescence measurements in the gas phase \cite{Catalan_JACS1993, Douhal_ChemPhys1996}. However, we note that the observed time constant is quite different than the previous gas-phase ultrafast spectroscopy measurement based on TRPES \cite{Lochbrunner_JChemPhys2001}, which reported 30 ps decay time constants even when using longer excitation wavelengths closer to the origin of the S$_0$ $\rightarrow$ S$_1$ transition. This shows the impact of the observable on the measurement of the kinetic time constants discussed in the introduction.

\begin{figure}
	\centering
	\includegraphics[width=\linewidth]{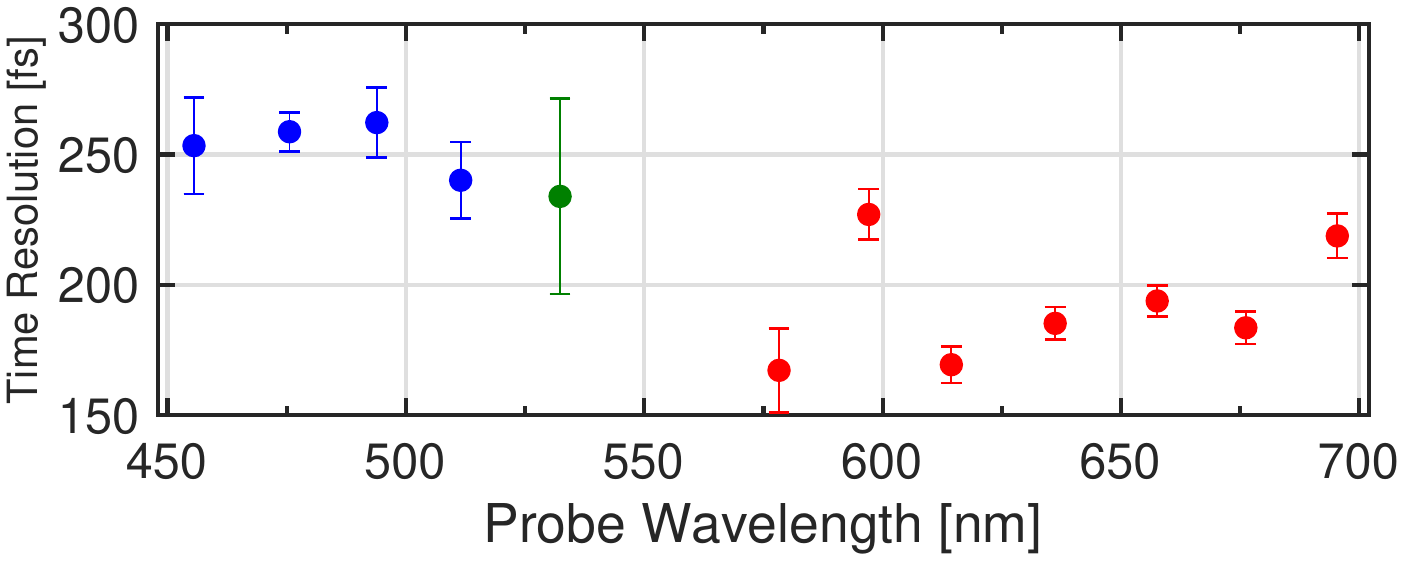}
	\caption{Spectrometer FWHM time resolution across the tuning range found by fitting the rising edge of HAN signal to an error function assuming instrument-limited response. Error bars are from fit.}
	\label{fig:sens_res}
\end{figure}

\ind With the assumption that the enol-keto tautomerization and the corresponding appearance of excited-state absorption and redshifted stimulated emission in HAN happen much faster than our time resolution \cite{Lochbrunner_JChemPhys2005}, we estimate the time resolution of the instrument by fitting the rising edge of the CE-TA traces with an error function. Figure \ref{fig:sens_res}a) shows the resulting extracted instrument response FWHM as a function of wavelength. Impulse response widths less than 275 fs are attained across the tuning range, with somewhat better time-resolution observed using the $2i$ comb.

\begin{figure}[t!]
	\centering
	\includegraphics[width=\linewidth]{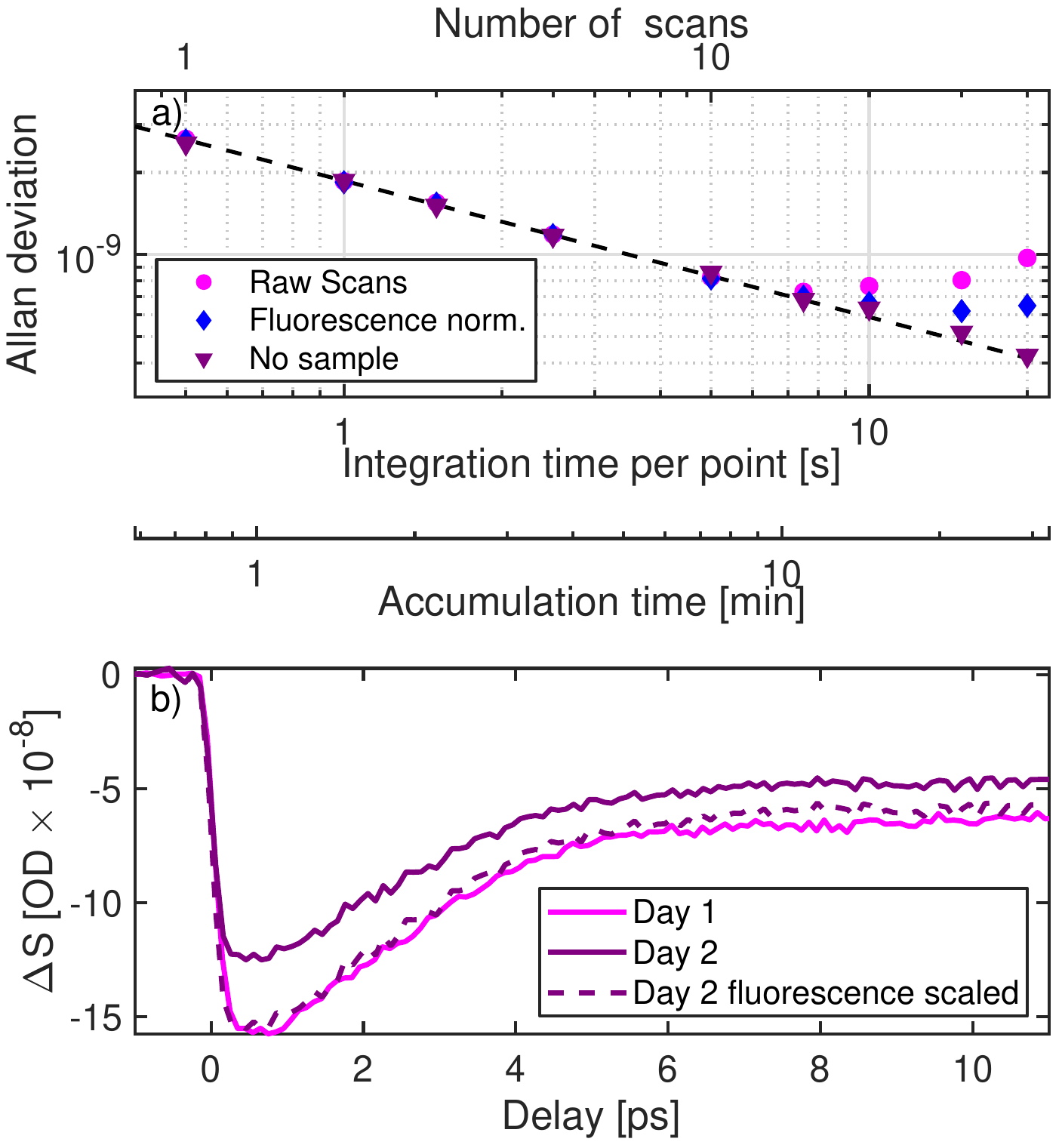}
	\caption{a) Allan deviation recorded using repetitive scans at 469 nm 0.5 s of integration per pump/probe delay. Without sample (triangles), the noise averages down with the inverse square root of the measurement time for as long as we have recorded data, following the dashed line with a slope of -1/2. With molecular signal, drift in the molecular column density on the $\sim$10 minute time scale causes the main limitation to averaging (circles), but this drift can be remedied to some extend using fluorescence normalization (diamonds). b) Two pump/probe traces taken days apart can largely be brought into coincidence using normalization to the fluorescence signal.}
	\label{fig:allan}
\end{figure}

\ind We now discuss the sensitivity of the instrument. There are two main sources of uncertainty (i.e. noise) to consider. The first is the optical noise floor of the system (figure \ref{fig:subtract}b) due to residual un-subtracted noise on the intracavity light and uncorrelated quantum noise in probe/reference detection. The second is drifts of the instrument over longer time scales required to assemble a full data set. Both can be quantified using an Allan deviation analysis \cite{Nagourney_Book2010, Allan_IEEE1966}. Figure \ref{fig:allan}a) shows the Allan deviation calculated from data sets where the same signals are scanned repeatedly.

\ind The intrinsic noise performance of the optical setup is captured by data taken without any sample (triangles on figure \ref{fig:allan}a). Without sample, the Allan deviation comes down with a slope of $-1/2$ on the log-log plot which indicates white-noise-limited performance (i.e. no drift). We observe this behavior for as long as we have averaged for and have seen noise down to  $\Delta$OD$ = 2.6 \times 10^{-11}$ (off the chart) after 90 minutes of integration without sample, with a corresponding normalized sensitivity of $\Delta\textrm{OD} = 2 \times 10^{-9} /\sqrt{\textrm{Hz}}$. Similar results with the 2i comb (not shown) give a sensitivity of $\Delta\textrm{OD} = 3 \times 10^{-9} /\sqrt{\textrm{Hz}}$. These results are consistent with the optical noise floor of the subtracted signal observed in figure \ref{fig:subtract} and comparable to the single-color result of Reber et al. \cite{Reber_Optica2016} of $\Delta\textrm{OD} = 1 \times 10^{-9} /\sqrt{\textrm{Hz}}$ \cite{Reber_Optica2016}, despite the significant additional complexity of the current setup, showing the intrinsic robustness of CE-TAS method.

\ind When accumulating an actual molecular signal, the uncertainty in $\Delta S$ (circles on figure \ref{fig:allan}a)) becomes dominated by drift for long accumulation times. These data are accumulated by repeated scanning of a pump/probe signal (parallel polarizations, 469 nm, 88 points, 0.5 s/point) such that the same pump/probe delay is re-encountered every 44 seconds. In this case, the Allan deviation differs (circles in \ref{fig:allan}a) from white-noise performance \cite{Nagourney_Book2010} and actually increases with averaging time for real accumulation times longer than 10 minutes. The main source of drift is variation in the sample molecular column density, which can be mitigated using the fluorescence monitor as we describe below.

\ind Figure \ref{fig:allan}b) demonstrates fluorescence normalization for an extreme case. The two scans, both at a probe wavelength of 455 nm and at equivalent sample backing pressures, were recorded several days apart. On the second day, there was less HAN remaining in the sample cell which resulted in reduced fluorescence and CE-TA signal. Scaling the day 2 data by the ratio of fluorescence signals brings the two TA signals back into coincidence. The Allan deviation of CE-TA data normalized using the fluorescence monitor is shown as diamonds on figure \ref{fig:allan}a)). With normalization, individual CE-TA pump/probe traces can be accumulated with a noise level of $\Delta$OD $ = 6 \times 10^{-10}$ with repetitive scans over a real accumulation time of 22 minutes. This corresponds to a S/N of 167 for this particular data set.

\subsection{CE-TAS Spectra}

\begin{figure}
	\centering
	\includegraphics[width=\linewidth]{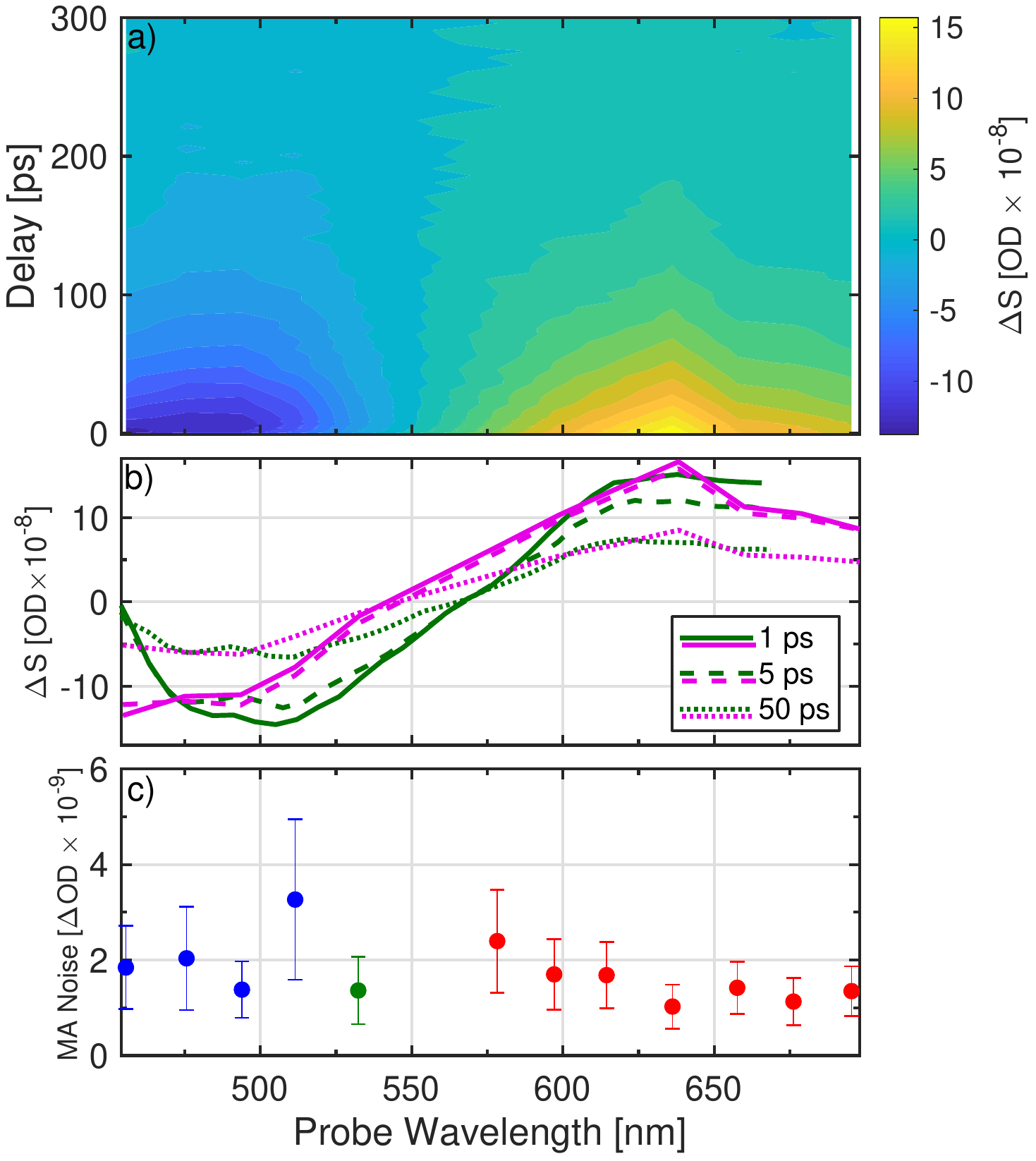}
	\caption{a) Magic-angle transient absorption map for jet-cooled HAN excited at 355 nm constructed from 12 probe wavelengths. Stimulated emission is observed on the blue side of the spectrum and excited state absorption on the red. b) Comparison of TA spectra from jet-cooled HAN (magenta) and HAN in cyclohexane (green) from reference \citenum{Lochbrunner_JChemPhys2005} at 1 ps (solid), 5 ps (long-dashed), and 50 ps (short-dashed) delays. The solution-phase data has been multiplied by one overall scale factor to make the comparison. c) Noise-levels attained as a function of wavelength for this full TA-map measurement.}
	\label{fig:HAN_contour}
\end{figure}

In figure \ref{fig:HAN_contour}a), we show a constructed magic angle transient absorption map for HAN in a He-seeded supersonic expansion (0.25 Bar stagnation pressure) working 3 mm from the nozzle. This spectrum is sampled at the same 12 discrete probe wavelengths as in figure \ref{fig:sens_res} and the same delay axis as figure \ref{fig:TAS_MA}c). For each wavelength, we take three scans for parallel and three scans for perpendicular pump/probe polarizations with an integration time of 1 s/delay. With 260 points/scan distributed over pump/probe delays out to 700 ps, each scan then takes 260 s = 4.3 minutes. Thus, we are accumulating data for a total of 26 minutes per wavelength. The entire spectrum comprising 12 wavelengths is collected over the course of a day. Figure \ref{fig:HAN_contour}c) shows the noise level for the magic angle signals, using fluorescence normalization, obtained under these practical conditions as a function of wavelength. 

\ind In figure \ref{fig:HAN_contour}b), we extract TA spectra of the molecule at 1, 5, and 50 ps delay and compare them to the TAS data reported by Lochbrunner et al. for HAN in cyclohexane \cite{Lochbrunner_JChemPhys2005}. The most obvious difference between our results from the jet-cooled molecule and the cyclohexane data is a solvatochromic blueshift of the TA data by $\sim$25 nm going from cyclohexane to gas-phase, similar (but not identical) to the 15 nm shift for fluorescence reported by Catalan et al. \cite{Catalan_JACS1993}. Furthermore, the differences in the TA spectra are not fully explained only by a solvatochromic shift. A more comprehensive analysis of the full HAN data is beyond the scope of this paper but will be the subject of a future publication including global analysis and comparison to \emph{ab initio} theory \cite{Mehmood_inprep2021}.

\subsection{Clusters}

\begin{figure}
	\centering
	\includegraphics[width=\linewidth]{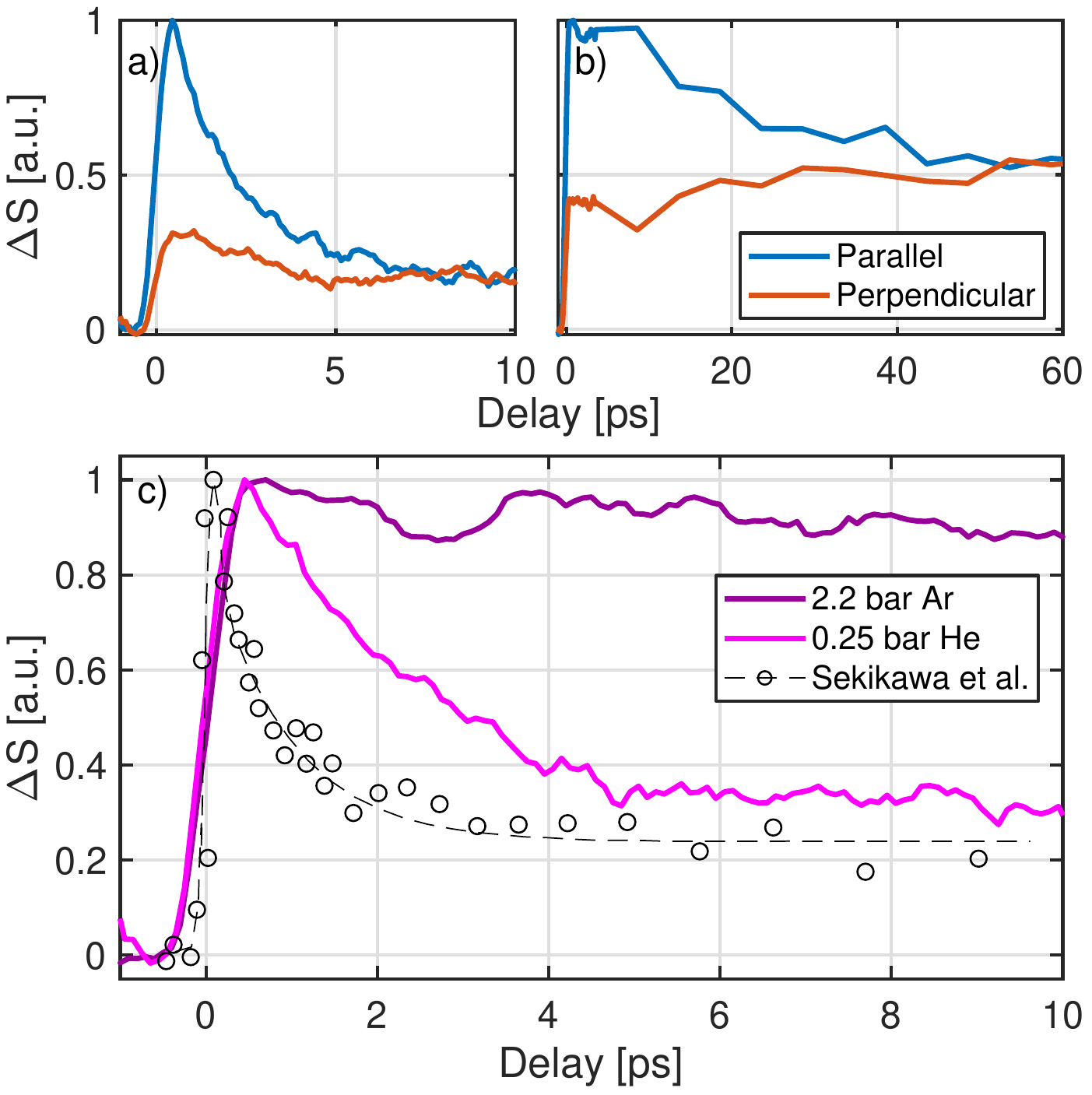}
	\caption{Parallel and perpendicular polarization CE-TAS data for SA excited at 355 nm in a 0.25 bar He expansion a) and 2.2 bar Ar expansion b). The polarization anisotropy transient decays much more slowly in the Ar data, indicating the formation of large Ar clusters. c) Magic angle data for the Ar expansion and He expansion compared to previous TRPES data recorded in jet-cooled SA from ref. \citenum{Sekikawa_JPCA2013}.}
	\label{fig:clusters}
\end{figure}

Working 3 mm from the 200 $\mu$m slit nozzle, we can easily generate clusters with sufficient column density for CE-TAS studies and we demonstrate this here. Figures \ref{fig:clusters}a) and b) shows an example of this for SA recorded at $\lambda_{\textss{probe}}$ = 455 nm expanded in helium and argon, respectively. Using He carrier gas, we observe the rotational anisotropy to decay in $\sim$10 ps, whereas for Ar carrier gas, the parallel and perpendicular polarizations data do not converge to the same signal until $\sim$50 ps. In the optimized ground state geometry calculated by Pijeau et al. \cite{Pijeau_JPCA2018}, the rotational constants of SA are $A$ = 0.066 cm$^{-1}$, $B$ = 0.0091 cm$^{-1}$, and $C$ = 0.0082 cm$^{-1}$, making the molecule nearly a symmetric top. For a symmetric top, the width of the rotational anisotropy transient scales as $1/\sqrt{B T}$\cite{Felker_JPhysChem1986}, indicating a large change in the rotational constant is required to explain the 5x increase in the width of the rotational anisotropy transient. It is important to note that for the case of SA, with much faster internal conversion than HAN and much smaller fluorescence yield of $\sim10^{-4}$\cite{Ziolek_PCCP2004}, the rotational anisotropy parameter $r'(t)$ constructed from the CE-TAS signals $\Delta S$ is free from the aforementioned complications due to reference subtraction and is bounded by $(-0.2,0.4)$. Assuming the temperature is similar in the two expansions, from the increased width of the rotational anisotropy, we estimate that the SA molecules have gained on average 24 Ar atoms, although actually this number should be taken as a lower bound since excitation of the molecule may promptly evaporate many Ar atoms, as commonly exploited in tag-loss spectroscopy \cite{Heine_IntRevPhysCHem2015}.

\ind The effect of Ar clustering on the internal dynamics of the molecule can be seen in the magic angle data shown in figure \ref{fig:clusters}c). For He expansions, where no clustering is expected, we observe fast decays of the TA signal in agreement with previous solution-phase TAS \cite{Ziolek_PCCP2004} and gas-phase TRPES  \cite{Sekikawa_JPCA2013}. However when forming large Ar clusters, the internal conversion is shut off and the excitation is long-lived, as shown in figure \ref{fig:clusters}. A corresponding large increase in the fluorescence signal is also observed, further supporting a suppression of internal conversion pathways in the Ar cluster. Also shown for direct comparison is TRPES data from Sekikawa et al. \cite{Sekikawa_JPCA2013}, which shows a much faster decay of the observable than TAS, similar to what we have also observed in HAN. More detailed analysis of the SA data, along with full spectra are the subject of a forthcoming paper \cite{Silfies_inprep2021}.

\section{Discussion}

In this article, we have described the performance of a broadband ultrasensitive spectrometer for recording transient absorption spectra with ultrafast time resolution. The overall performance of the spectrometer is comparable to a previous 1-color demonstration of the main concept in molecular I$_2$ \cite{Reber_Optica2016}, despite the significant additional complexity of both the optical setup and molecular beam system necessary to go past demonstrations and record data on chemically relevant systems. We have also demonstrated the linkages shown in figure \ref{fig:punnett} by directly comparing cavity-enhanced transient absorption data to solution-phase TA measurements and gas-phase TRPES for two example systems. We expect a wealth of information can be extracted from such comparisons going forward, given the large body of high-quality data existing from these well-established techniques. Furthermore, performing CE-TAS measurement on clusters can enable a detailed microscopic understanding of the effect of the solvent on molecular dynamics, as has been done for linear spectroscopy.

\ind For electronically excited molecules, UV-VIS CE-TAS offers a complimentary ultrafast observable to those provided by well-established TRPES methods, with the idea that via comparison to theory more information can be extracted from the combination than can be had from either observable alone. This multi-observable approach has recently been promoted by others for the combination of diffraction and spectroscopy data \cite{Liu_PRX2020}.

\ind The methods described here can also be implemented in the mid-infrared to study purely vibrational dynamics on the electronic ground state, and we are actively working on developing cavity-enhanced two-dimensional infrared spectroscopy (CE-2DIR) \cite{Allison_JPhysB2017}. It is important to note that in contrast to the current work, which provides a complimentary view of the dynamics of gas-phase molecules after electronic excitation, for which other action-based spectroscopy methods exist, an action-based analog of 2DIR with ultrafast time resolution does not currently exist. In many ways, we expect CE-2DIR spectroscopy to be less technically challenging than the current demonstration due to the reduced bandwidth requirements of 2DIR and also less difficulties with mirror contamination due to the absence of UV light cracking residual hydrocarbons in the vacuum system. 

\ind Finally, we note that the methods demonstrated here can be adapted to solids, liquids, and sparsely covered surfaces, as has been done for cavity-enhanced linear spectroscopy \cite{Vallance_bookchapter2014}. For example, inclusion of a reflection off a glass/liquid interface into the cavity could be used to perform cavity-enhanced ultrafast attenuated total reflectance spectroscopy on molecules at the interface. Translating the current sensitivity to a molecular film indicates that coverages below 10$^{-4}$ monolayer could be investigated. The challenges in adapting CE-TAS to condensed-phase contexts are 1) managing the dispersion and loss of additional intracavity elements and 2) managing sample excitation and refresh rate. While it is likely that compromises regarding 1) and 2) would reduce performance, CE-TAS methods could still find applicability for small-signal condensed phase measurements inaccessible with other techniques.

\section{Acknowledgements}

This work was supported by the U. S. National Science Foundation under award number 1708743 and the U. S. Air Force Office of Scientific Research under grant number FA9550-20-1-0259. M. C. Silfies acknowledges support from the GAANN program of the U. S. Dept. of Education. G. Kowzan acknowledges support from the National Science Centre, Poland scholarship 2017/24/T/ST2/00242. The authors thank S.A. Diddams, H. Timmers, A. Kowligy, N. Nader, and G. Ycas for assistance with developing the dispersive-wave-shifted Er:fiber comb.



\balance


\providecommand*{\mcitethebibliography}{\thebibliography}
\csname @ifundefined\endcsname{endmcitethebibliography}
{\let\endmcitethebibliography\endthebibliography}{}

\bibliographystyle{rsc} 

\begin{mcitethebibliography}{44}
	\providecommand*{\natexlab}[1]{#1}
	\providecommand*{\mciteSetBstSublistMode}[1]{}
	\providecommand*{\mciteSetBstMaxWidthForm}[2]{}
	\providecommand*{\mciteBstWouldAddEndPuncttrue}
	{\def\EndOfBibitem{\unskip.}}
	\providecommand*{\mciteBstWouldAddEndPunctfalse}
	{\let\EndOfBibitem\relax}
	\providecommand*{\mciteSetBstMidEndSepPunct}[3]{}
	\providecommand*{\mciteSetBstSublistLabelBeginEnd}[3]{}
	\providecommand*{\EndOfBibitem}{}
	\mciteSetBstSublistMode{f}
	\mciteSetBstMaxWidthForm{subitem}
	{(\emph{\alph{mcitesubitemcount}})}
	\mciteSetBstSublistLabelBeginEnd{\mcitemaxwidthsubitemform\space}
	{\relax}{\relax}
	
	\bibitem[van Stokkum \emph{et~al.}(2004)van Stokkum, Larsen, and van
	Grondelle]{vanStokkum_GlobalAnalysis2004}
	I.~H. van Stokkum, D.~S. Larsen and R.~van Grondelle, \emph{Biochimica et
		Biophysica Acta (BBA) - Bioenergetics}, 2004, \textbf{1657}, 82 -- 104\relax
	\mciteBstWouldAddEndPuncttrue
	\mciteSetBstMidEndSepPunct{\mcitedefaultmidpunct}
	{\mcitedefaultendpunct}{\mcitedefaultseppunct}\relax
	\EndOfBibitem
	\bibitem[Hudock \emph{et~al.}(2007)Hudock, Levine, Thompson, Satzger, Townsend,
	Gador, Ullrich, Stolow, and Mart{\'\i}nez]{Hudock_JPCA2007}
	H.~R. Hudock, B.~G. Levine, A.~L. Thompson, H.~Satzger, D.~Townsend, N.~Gador,
	S.~Ullrich, A.~Stolow and T.~J. Mart{\'\i}nez, \emph{The Journal of Physical
		Chemistry A}, 2007, \textbf{111}, 8500--8508\relax
	\mciteBstWouldAddEndPuncttrue
	\mciteSetBstMidEndSepPunct{\mcitedefaultmidpunct}
	{\mcitedefaultendpunct}{\mcitedefaultseppunct}\relax
	\EndOfBibitem
	\bibitem[Tao \emph{et~al.}(2011)Tao, Allison, Wright, Stooke, Khurmi, van
	Tilborg, Liu, Falcone, Belkacem, and Martinez]{Tao_JChemPhys2011}
	H.~Tao, T.~K. Allison, T.~W. Wright, A.~M. Stooke, C.~Khurmi, J.~van Tilborg,
	Y.~Liu, R.~W. Falcone, A.~Belkacem and T.~J. Martinez, \emph{The Journal of
		Chemical Physics}, 2011, \textbf{134}, 244306\relax
	\mciteBstWouldAddEndPuncttrue
	\mciteSetBstMidEndSepPunct{\mcitedefaultmidpunct}
	{\mcitedefaultendpunct}{\mcitedefaultseppunct}\relax
	\EndOfBibitem
	\bibitem[Adachi \emph{et~al.}(2019)Adachi, Schatteburg, Humeniuk, Mitri{\'c},
	and Suzuki]{Adachi_PCCP2019}
	S.~Adachi, T.~Schatteburg, A.~Humeniuk, R.~Mitri{\'c} and T.~Suzuki,
	\emph{Phys. Chem. Chem. Phys.}, 2019, \textbf{21}, 13902--13905\relax
	\mciteBstWouldAddEndPuncttrue
	\mciteSetBstMidEndSepPunct{\mcitedefaultmidpunct}
	{\mcitedefaultendpunct}{\mcitedefaultseppunct}\relax
	\EndOfBibitem
	\bibitem[Liu \emph{et~al.}(2020)Liu, Horton, Yang, Nunes, Shen, Wolf, Forbes,
	Cheng, Moore, Centurion, Hegazy, Li, Lin, Stolow, Hockett, Rozgonyi,
	Marquetand, Wang, and Weinacht]{Liu_PRX2020}
	Y.~Liu, S.~L. Horton, J.~Yang, J.~P.~F. Nunes, X.~Shen, T.~J.~A. Wolf,
	R.~Forbes, C.~Cheng, B.~Moore, M.~Centurion, K.~Hegazy, R.~Li, M.-F. Lin,
	A.~Stolow, P.~Hockett, T.~Rozgonyi, P.~Marquetand, X.~Wang and T.~Weinacht,
	\emph{Phys. Rev. X}, 2020, \textbf{10}, 021016\relax
	\mciteBstWouldAddEndPuncttrue
	\mciteSetBstMidEndSepPunct{\mcitedefaultmidpunct}
	{\mcitedefaultendpunct}{\mcitedefaultseppunct}\relax
	\EndOfBibitem
	\bibitem[Allison \emph{et~al.}(2012)Allison, Tao, Glover, Wright, Stooke,
	Khurmi, van Tilborg, Liu, Falcone, Martinez, and
	Belkacem]{Allison_JChemPhys2012}
	T.~K. Allison, H.~Tao, W.~J. Glover, T.~W. Wright, A.~M. Stooke, C.~Khurmi,
	J.~van Tilborg, Y.~Liu, R.~W. Falcone, T.~J. Martinez and A.~Belkacem,
	\emph{The Journal of Chemical Physics}, 2012, \textbf{136}, 124317\relax
	\mciteBstWouldAddEndPuncttrue
	\mciteSetBstMidEndSepPunct{\mcitedefaultmidpunct}
	{\mcitedefaultendpunct}{\mcitedefaultseppunct}\relax
	\EndOfBibitem
	\bibitem[Schuurman and Stolow(2018)]{Schuurman_AnnRevPhysChem2018}
	M.~S. Schuurman and A.~Stolow, \emph{Annual Review of Physical Chemistry},
	2018, \textbf{69}, 427--450\relax
	\mciteBstWouldAddEndPuncttrue
	\mciteSetBstMidEndSepPunct{\mcitedefaultmidpunct}
	{\mcitedefaultendpunct}{\mcitedefaultseppunct}\relax
	\EndOfBibitem
	\bibitem[Stolow \emph{et~al.}(2004)Stolow, Bragg, and
	Neumark]{Stolow_ChemRev2004}
	A.~Stolow, A.~E. Bragg and D.~M. Neumark, \emph{Chemical Reviews}, 2004,
	\textbf{104}, 1719--1758\relax
	\mciteBstWouldAddEndPuncttrue
	\mciteSetBstMidEndSepPunct{\mcitedefaultmidpunct}
	{\mcitedefaultendpunct}{\mcitedefaultseppunct}\relax
	\EndOfBibitem
	\bibitem[Middleton \emph{et~al.}(2009)Middleton, de~La~Harpe, Su, Law,
	Crespo-Hern{\'a}ndez, and Kohler]{Kohler_AnnRevPhysChem2009}
	C.~T. Middleton, K.~de~La~Harpe, C.~Su, Y.~K. Law, C.~E. Crespo-Hern{\'a}ndez
	and B.~Kohler, \emph{Annual Review of Physical Chemistry}, 2009, \textbf{60},
	217--239\relax
	\mciteBstWouldAddEndPuncttrue
	\mciteSetBstMidEndSepPunct{\mcitedefaultmidpunct}
	{\mcitedefaultendpunct}{\mcitedefaultseppunct}\relax
	\EndOfBibitem
	\bibitem[Saigusa(2006)]{Saigusa_JPhotoChemBio2007}
	H.~Saigusa, \emph{Journal of Photochemistry and Photobiology C: Photochemistry
		Reviews}, 2006, \textbf{7}, 197 -- 210\relax
	\mciteBstWouldAddEndPuncttrue
	\mciteSetBstMidEndSepPunct{\mcitedefaultmidpunct}
	{\mcitedefaultendpunct}{\mcitedefaultseppunct}\relax
	\EndOfBibitem
	\bibitem[Ullrich \emph{et~al.}(2004)Ullrich, Schultz, Zgierski, and
	Stolow]{Ullrich_PCCP2004}
	S.~Ullrich, T.~Schultz, M.~Z. Zgierski and A.~Stolow, \emph{Phys. Chem. Chem.
		Phys.}, 2004, \textbf{6}, 2796--2801\relax
	\mciteBstWouldAddEndPuncttrue
	\mciteSetBstMidEndSepPunct{\mcitedefaultmidpunct}
	{\mcitedefaultendpunct}{\mcitedefaultseppunct}\relax
	\EndOfBibitem
	\bibitem[Tseng \emph{et~al.}(2012)Tseng, S{\'a}ndor, Kotur, Weinacht, and
	Matsika]{Tseng_JPCA2012}
	C.-h. Tseng, P.~S{\'a}ndor, M.~Kotur, T.~C. Weinacht and S.~Matsika, \emph{The
		Journal of Physical Chemistry A}, 2012, \textbf{116}, 2654--2661\relax
	\mciteBstWouldAddEndPuncttrue
	\mciteSetBstMidEndSepPunct{\mcitedefaultmidpunct}
	{\mcitedefaultendpunct}{\mcitedefaultseppunct}\relax
	\EndOfBibitem
	\bibitem[Reber \emph{et~al.}(2016)Reber, Chen, and Allison]{Reber_Optica2016}
	M.~A.~R. Reber, Y.~Chen and T.~K. Allison, \emph{Optica}, 2016, \textbf{3},
	311--317\relax
	\mciteBstWouldAddEndPuncttrue
	\mciteSetBstMidEndSepPunct{\mcitedefaultmidpunct}
	{\mcitedefaultendpunct}{\mcitedefaultseppunct}\relax
	\EndOfBibitem
	\bibitem[Schriever \emph{et~al.}(2008)Schriever, Lochbrunner, Riedle, and
	Nesbitt]{Schriever_RSI2008}
	C.~Schriever, S.~Lochbrunner, E.~Riedle and D.~J. Nesbitt, \emph{Rev. Sci.
		Inst.}, 2008, \textbf{79}, 013107\relax
	\mciteBstWouldAddEndPuncttrue
	\mciteSetBstMidEndSepPunct{\mcitedefaultmidpunct}
	{\mcitedefaultendpunct}{\mcitedefaultseppunct}\relax
	\EndOfBibitem
	\bibitem[Chen \emph{et~al.}(2019)Chen, Silfies, Kowzan, Bautista, and
	Allison]{Chen_ApplPhysB2019}
	Y.~Chen, M.~C. Silfies, G.~Kowzan, J.~M. Bautista and T.~K. Allison,
	\emph{Applied Physics B}, 2019, \textbf{125}, 81\relax
	\mciteBstWouldAddEndPuncttrue
	\mciteSetBstMidEndSepPunct{\mcitedefaultmidpunct}
	{\mcitedefaultendpunct}{\mcitedefaultseppunct}\relax
	\EndOfBibitem
	\bibitem[Silfies \emph{et~al.}(2020)Silfies, Kowzan, Chen, Lewis, Hou, Baehre,
	Gross, and Allison]{Silfies_OptLett2020}
	M.~C. Silfies, G.~Kowzan, Y.~Chen, N.~Lewis, R.~Hou, R.~Baehre, T.~Gross and
	T.~K. Allison, \emph{Opt. Lett.}, 2020, \textbf{45}, 2123--2126\relax
	\mciteBstWouldAddEndPuncttrue
	\mciteSetBstMidEndSepPunct{\mcitedefaultmidpunct}
	{\mcitedefaultendpunct}{\mcitedefaultseppunct}\relax
	\EndOfBibitem
	\bibitem[Suzuki(2019)]{Suzuki_JChemPhys2019}
	T.~Suzuki, \emph{The Journal of Chemical Physics}, 2019, \textbf{151},
	090901\relax
	\mciteBstWouldAddEndPuncttrue
	\mciteSetBstMidEndSepPunct{\mcitedefaultmidpunct}
	{\mcitedefaultendpunct}{\mcitedefaultseppunct}\relax
	\EndOfBibitem
	\bibitem[Maser \emph{et~al.}(2017)Maser, Ycas, Depetri, Cruz, and
	Diddams]{Maser_ApplPhysB2017}
	D.~L. Maser, G.~Ycas, W.~I. Depetri, F.~C. Cruz and S.~A. Diddams,
	\emph{Applied Physics B}, 2017, \textbf{123}, 142\relax
	\mciteBstWouldAddEndPuncttrue
	\mciteSetBstMidEndSepPunct{\mcitedefaultmidpunct}
	{\mcitedefaultendpunct}{\mcitedefaultseppunct}\relax
	\EndOfBibitem
	\bibitem[Li \emph{et~al.}(2016)Li, Reber, Corder, Chen, Zhao, and
	Allison]{Li_RSI2016}
	X.~Li, M.~A.~R. Reber, C.~Corder, Y.~Chen, P.~Zhao and T.~K. Allison,
	\emph{Review of Scientific Instruments}, 2016, \textbf{87}, 093114\relax
	\mciteBstWouldAddEndPuncttrue
	\mciteSetBstMidEndSepPunct{\mcitedefaultmidpunct}
	{\mcitedefaultendpunct}{\mcitedefaultseppunct}\relax
	\EndOfBibitem
	\bibitem[Jones and Ye(2004)]{Jones_OptLett2004}
	R.~J. Jones and J.~Ye, \emph{Opt. Lett.}, 2004, \textbf{29}, 2812--2814\relax
	\mciteBstWouldAddEndPuncttrue
	\mciteSetBstMidEndSepPunct{\mcitedefaultmidpunct}
	{\mcitedefaultendpunct}{\mcitedefaultseppunct}\relax
	\EndOfBibitem
	\bibitem[Jones and Ye(2002)]{Jones_OptLett2002}
	R.~J. Jones and J.~Ye, \emph{Opt. Lett.}, 2002, \textbf{27}, 1848--1850\relax
	\mciteBstWouldAddEndPuncttrue
	\mciteSetBstMidEndSepPunct{\mcitedefaultmidpunct}
	{\mcitedefaultendpunct}{\mcitedefaultseppunct}\relax
	\EndOfBibitem
	\bibitem[Nagourney(2010)]{Nagourney_Book2010}
	W.~Nagourney, \emph{Quantum Electronics for Atomic Physics}, Oxford University
	Press, 2010\relax
	\mciteBstWouldAddEndPuncttrue
	\mciteSetBstMidEndSepPunct{\mcitedefaultmidpunct}
	{\mcitedefaultendpunct}{\mcitedefaultseppunct}\relax
	\EndOfBibitem
	\bibitem[Siegman(1986)]{Siegman_book1986}
	A.~Siegman, \emph{Lasers}, University Science Books, 1986\relax
	\mciteBstWouldAddEndPuncttrue
	\mciteSetBstMidEndSepPunct{\mcitedefaultmidpunct}
	{\mcitedefaultendpunct}{\mcitedefaultseppunct}\relax
	\EndOfBibitem
	\bibitem[Allison(2017)]{Allison_JPhysB2017}
	T.~K. Allison, \emph{Journal of Physics B: Atomic, Molecular and Optical
		Physics}, 2017, \textbf{50}, 044004\relax
	\mciteBstWouldAddEndPuncttrue
	\mciteSetBstMidEndSepPunct{\mcitedefaultmidpunct}
	{\mcitedefaultendpunct}{\mcitedefaultseppunct}\relax
	\EndOfBibitem
	\bibitem[Miller(1988)]{Miller:1988}
	D.~Miller, in \emph{Atomic and Molecular Beam Methods}, ed. G.~Scoles, Oxford
	University Press, 1988\relax
	\mciteBstWouldAddEndPuncttrue
	\mciteSetBstMidEndSepPunct{\mcitedefaultmidpunct}
	{\mcitedefaultendpunct}{\mcitedefaultseppunct}\relax
	\EndOfBibitem
	\bibitem[Jarzeba \emph{et~al.}(2002)Jarzeba, Matylitsky, Weichert, and
	Riehn]{Jarzeba_PCCP2002}
	W.~Jarzeba, V.~V. Matylitsky, A.~Weichert and C.~Riehn, \emph{Phys. Chem. Chem.
		Phys.}, 2002, \textbf{4}, 451--454\relax
	\mciteBstWouldAddEndPuncttrue
	\mciteSetBstMidEndSepPunct{\mcitedefaultmidpunct}
	{\mcitedefaultendpunct}{\mcitedefaultseppunct}\relax
	\EndOfBibitem
	\bibitem[Pupeza \emph{et~al.}(2010)Pupeza, Gu, Fill, Eidam, Limpert,
	T\"{u}nnermann, Krausz, and Udem]{Pupeza_OptExp2010}
	I.~Pupeza, X.~Gu, E.~Fill, T.~Eidam, J.~Limpert, A.~T\"{u}nnermann, F.~Krausz
	and T.~Udem, \emph{Opt. Express}, 2010, \textbf{18}, 26184--26195\relax
	\mciteBstWouldAddEndPuncttrue
	\mciteSetBstMidEndSepPunct{\mcitedefaultmidpunct}
	{\mcitedefaultendpunct}{\mcitedefaultseppunct}\relax
	\EndOfBibitem
	\bibitem[Hobbs(1997)]{Hobbs_AppOpt1997}
	P.~C.~D. Hobbs, \emph{Appl. Opt.}, 1997, \textbf{36}, 903--920\relax
	\mciteBstWouldAddEndPuncttrue
	\mciteSetBstMidEndSepPunct{\mcitedefaultmidpunct}
	{\mcitedefaultendpunct}{\mcitedefaultseppunct}\relax
	\EndOfBibitem
	\bibitem[Gagliardi and Loock(2013)]{Gagliardi_Book2013}
	\emph{Cavity Enhanced Spectroscopy and Sensing}, ed. G.~Gagliardi and H.-P.
	Loock, Springer, 2013\relax
	\mciteBstWouldAddEndPuncttrue
	\mciteSetBstMidEndSepPunct{\mcitedefaultmidpunct}
	{\mcitedefaultendpunct}{\mcitedefaultseppunct}\relax
	\EndOfBibitem
	\bibitem[Weiner(2009)]{Weiner_book2009}
	A.~Weiner, \emph{Ultrafast Optics}, Wiley, 2009\relax
	\mciteBstWouldAddEndPuncttrue
	\mciteSetBstMidEndSepPunct{\mcitedefaultmidpunct}
	{\mcitedefaultendpunct}{\mcitedefaultseppunct}\relax
	\EndOfBibitem
	\bibitem[Felker and Zewail(1987)]{Felker_JChemPhys1987}
	P.~M. Felker and A.~H. Zewail, \emph{The Journal of Chemical Physics}, 1987,
	\textbf{86}, 2460--2482\relax
	\mciteBstWouldAddEndPuncttrue
	\mciteSetBstMidEndSepPunct{\mcitedefaultmidpunct}
	{\mcitedefaultendpunct}{\mcitedefaultseppunct}\relax
	\EndOfBibitem
	\bibitem[Lochbrunner \emph{et~al.}(2005)Lochbrunner, Szeghalmi, Stock, and
	Schmitt]{Lochbrunner_JChemPhys2005}
	S.~Lochbrunner, A.~Szeghalmi, K.~Stock and M.~Schmitt, \emph{The Journal of
		Chemical Physics}, 2005, \textbf{122}, 244315\relax
	\mciteBstWouldAddEndPuncttrue
	\mciteSetBstMidEndSepPunct{\mcitedefaultmidpunct}
	{\mcitedefaultendpunct}{\mcitedefaultseppunct}\relax
	\EndOfBibitem
	\bibitem[Zi{\'o}{\l}ek \emph{et~al.}(2004)Zi{\'o}{\l}ek, Kubicki, Maciejewski,
	Naskrecki, and Grabowska]{Ziolek_PCCP2004}
	M.~Zi{\'o}{\l}ek, J.~Kubicki, A.~Maciejewski, R.~Naskrecki and A.~Grabowska,
	\emph{Phys. Chem. Chem. Phys.}, 2004, \textbf{6}, 4682--4689\relax
	\mciteBstWouldAddEndPuncttrue
	\mciteSetBstMidEndSepPunct{\mcitedefaultmidpunct}
	{\mcitedefaultendpunct}{\mcitedefaultseppunct}\relax
	\EndOfBibitem
	\bibitem[Lochbrunner \emph{et~al.}(2001)Lochbrunner, Schultz, Schmitt, Shaffer,
	Zgierski, and Stolow]{Lochbrunner_JChemPhys2001}
	S.~Lochbrunner, T.~Schultz, M.~Schmitt, J.~P. Shaffer, M.~Z. Zgierski and
	A.~Stolow, \emph{The Journal of Chemical Physics}, 2001, \textbf{114},
	2519--2522\relax
	\mciteBstWouldAddEndPuncttrue
	\mciteSetBstMidEndSepPunct{\mcitedefaultmidpunct}
	{\mcitedefaultendpunct}{\mcitedefaultseppunct}\relax
	\EndOfBibitem
	\bibitem[Sekikawa \emph{et~al.}(2013)Sekikawa, Schalk, Wu, Boguslavskiy, and
	Stolow]{Sekikawa_JPCA2013}
	T.~Sekikawa, O.~Schalk, G.~Wu, A.~E. Boguslavskiy and A.~Stolow, \emph{The
		Journal of Physical Chemistry A}, 2013, \textbf{117}, 2971--2979\relax
	\mciteBstWouldAddEndPuncttrue
	\mciteSetBstMidEndSepPunct{\mcitedefaultmidpunct}
	{\mcitedefaultendpunct}{\mcitedefaultseppunct}\relax
	\EndOfBibitem
	\bibitem[Catalan and del Valle(1993)]{Catalan_JACS1993}
	J.~Catalan and J.~C. del Valle, \emph{Journal of the American Chemical
		Society}, 1993, \textbf{115}, 4321--4325\relax
	\mciteBstWouldAddEndPuncttrue
	\mciteSetBstMidEndSepPunct{\mcitedefaultmidpunct}
	{\mcitedefaultendpunct}{\mcitedefaultseppunct}\relax
	\EndOfBibitem
	\bibitem[Douhal \emph{et~al.}(1996)Douhal, Lahmani, and
	Zewail]{Douhal_ChemPhys1996}
	A.~Douhal, F.~Lahmani and A.~H. Zewail, \emph{Chemical Physics}, 1996,
	\textbf{207}, 477--498\relax
	\mciteBstWouldAddEndPuncttrue
	\mciteSetBstMidEndSepPunct{\mcitedefaultmidpunct}
	{\mcitedefaultendpunct}{\mcitedefaultseppunct}\relax
	\EndOfBibitem
	\bibitem[{Allan}(1966)]{Allan_IEEE1966}
	D.~W. {Allan}, \emph{Proceedings of the IEEE}, 1966, \textbf{54},
	221--230\relax
	\mciteBstWouldAddEndPuncttrue
	\mciteSetBstMidEndSepPunct{\mcitedefaultmidpunct}
	{\mcitedefaultendpunct}{\mcitedefaultseppunct}\relax
	\EndOfBibitem
	\bibitem[Mehmood \emph{et~al.}(2021)Mehmood, Silfies, Kowzan, Lewis, Allison,
	and Levine]{Mehmood_inprep2021}
	A.~Mehmood, M.~C. Silfies, G.~Kowzan, N.~Lewis, T.~K. Allison and B.~G. Levine,
	\emph{in preparation}, 2021\relax
	\mciteBstWouldAddEndPuncttrue
	\mciteSetBstMidEndSepPunct{\mcitedefaultmidpunct}
	{\mcitedefaultendpunct}{\mcitedefaultseppunct}\relax
	\EndOfBibitem
	\bibitem[Pijeau \emph{et~al.}(2018)Pijeau, Foster, and
	Hohenstein]{Pijeau_JPCA2018}
	S.~Pijeau, D.~Foster and E.~G. Hohenstein, \emph{The Journal of Physical
		Chemistry A}, 2018, \textbf{122}, 5555--5562\relax
	\mciteBstWouldAddEndPuncttrue
	\mciteSetBstMidEndSepPunct{\mcitedefaultmidpunct}
	{\mcitedefaultendpunct}{\mcitedefaultseppunct}\relax
	\EndOfBibitem
	\bibitem[Felker \emph{et~al.}(1986)Felker, Baskin, and
	Zewail]{Felker_JPhysChem1986}
	P.~M. Felker, J.~S. Baskin and A.~H. Zewail, \emph{The Journal of Physical
		Chemistry}, 1986, \textbf{90}, 724--728\relax
	\mciteBstWouldAddEndPuncttrue
	\mciteSetBstMidEndSepPunct{\mcitedefaultmidpunct}
	{\mcitedefaultendpunct}{\mcitedefaultseppunct}\relax
	\EndOfBibitem
	\bibitem[Heine and Asmis(2015)]{Heine_IntRevPhysCHem2015}
	N.~Heine and K.~R. Asmis, \emph{International Reviews in Physical Chemistry},
	2015, \textbf{34}, 1--34\relax
	\mciteBstWouldAddEndPuncttrue
	\mciteSetBstMidEndSepPunct{\mcitedefaultmidpunct}
	{\mcitedefaultendpunct}{\mcitedefaultseppunct}\relax
	\EndOfBibitem
	\bibitem[Silfies and et~al.(2021)]{Silfies_inprep2021}
	M.~Silfies and et~al., \emph{in preparation}, 2021\relax
	\mciteBstWouldAddEndPuncttrue
	\mciteSetBstMidEndSepPunct{\mcitedefaultmidpunct}
	{\mcitedefaultendpunct}{\mcitedefaultseppunct}\relax
	\EndOfBibitem
	\bibitem[Vallance and Rushworth(2014)]{Vallance_bookchapter2014}
	C.~Vallance and C.~M. Rushworth, in \emph{Cavity-Enhanced Spectroscopy and
		Sensing}, Springer-Verlag, 2014\relax
	\mciteBstWouldAddEndPuncttrue
	\mciteSetBstMidEndSepPunct{\mcitedefaultmidpunct}
	{\mcitedefaultendpunct}{\mcitedefaultseppunct}\relax
	\EndOfBibitem
\end{mcitethebibliography}

\end{document}